\documentclass[letter,oneside,9pt,twocolumn,article]{aiaa}
\usepackage{graphicx}
\usepackage{rotating}
\usepackage{bm}
\usepackage{mathtools}
\usepackage[bf,BF,small]{subfigure}
\usepackage{adjustbox}
\usepackage{tabularx}
\usepackage{multirow}
\usepackage[hidelinks]{hyperref} 
\usepackage{float}
\usepackage{xcolor}
\usepackage{mwe}
\usepackage[markcase=noupper]{scrlayer-scrpage}
\usepackage{flushend}
\ihead*{{\it AIAA Journal}, Vol.~63, No.~8, 2025, pp.~3035-3047, \href{http://doi.org/10.2514/1.J064125}{doi:~10.2514/1.J064125}}
\ohead*{\pagemark}
\ifoot*{\rm }% left side of footer
\ofoot*{\rm Unrestricted content}% right side of footer
\newcommand{\alb}{\vspace{0.1cm}\\} % array line break

\renewcommand{\vec}[1]{\bm{#1}}

\newcommand{\ns}{{{n}_{\rm s}}}

\renewcommand{\fontsizetable}{\footnotesize\scalefont{1.0}}

\renewcommand{\vec}[1]{\bm{#1}}
\newcommand\blfootnote[1]{%
  \begingroup
  \renewcommand\thefootnote{}\footnote{#1}%
  \addtocounter{footnote}{-1}%
  \endgroup
}

\author{ Bernard Parent\thanks{Associate Professor, Aerospace and Mechanical Engineering, Associate Fellow AIAA, bparent@arizona.edu}
~~and~ 
Felipe Martin Rodriguez Fuentes \thanks{Graduate Student, Aerospace and Mechanical Engineering} 
~~and~
Spencer LaFoley \thanks{Graduate Student, Aerospace and Mechanical Engineering} 
\\
          {\it University of Arizona, Tucson, AZ 85721}\\
    }

\title{Electrodeless Magnetohydrodynamic Local Force Generator\\ for Aerocapture}

\abstract{
This paper presents a novel magnetohydrodynamics (MHD) system for planetary entry aerocapture. The system is advantaged over previous approaches by having the following two characteristics: (i) it can be deployed locally to one or various flow regions, and (ii) it does not make use of electrodes. Previous MHD systems for planetary entry were either electrodeless global systems or two-electrode local systems. The proposed novel MHD system employs two magnets to establish a current loop resulting in a Faraday electromotive force (EMF). The first magnet is positioned to ensure the magnetic field faces outward from the shell, while the second magnet is oriented to ensure the magnetic field faces inward toward the shell. Preliminary findings demonstrate that when located on the surface of an Earth entry capsule at a flight Mach number of 35, the novel electrodeless MHD system can generate forces several times greater than a two-electrode system while utilizing the same magnetic field strength. The study is conducted entirely through numerical simulation using CFDWARP, a computational fluid dynamics (CFD) code that employs advanced numerical methods allowing for the full coupling between aerodynamics, magnetohydrodynamics, and non-neutral plasma sheaths.  The physical model includes an 11-species finite-rate chemical solver including real gas effects, the drift-diffusion model for all charged species,  along with an electric field potential equation that satisfies Gauss's law.
}

\nomenclature{
  \begin{nomenclaturelist}{Roman symbols}
   \item[$A$] coefficient used to determine reaction rates in Arrhenius form, $\textrm{cm}^{3b}\cdot \textrm{mole}^{-b} \cdot \textrm{s}^{-1} \cdot \textrm{K}^{-n}$ where $b=0$ for one-body reactions, $b=1$ for two-body reactions and $b=2$ for three-body reactions
   \item[$A_0$] constant in Richardson's emission law, $\mathrm{A/m^2K^2}$
   \item[$\vec{B}$] net magnetic field vector {(sum of applied and induced magnetic fields)}, T
   {\item[$\vec{B}_{\rm applied}$] applied magnetic field vector, T
   \item[$\vec{B}_{\rm induced}$] induced magnetic field vector, T}
   \item[$C_k$] charge of $k$th species, C
   \item[$D$] computational domain depth, m
   \item[$e_{k}$] specific energy of $k$th species, $h_k-P_k/\rho_k$
   \item[$e_{\rm t}$] total specific energy, J/kg
   \item[$E$] reaction activation energy, cal/mol
   \item[$\vec{E}$] electric field vector, V/m
   \item[$E_i$] electric field along $i$th dimension, V/m
   \item[$E^\star$] reduced electric field $|\vec{E}|/N$, V$\mathrm{m^2}$
   \item[$\cal E$] Faraday electromotive force (EMF), V
   \item[$F$] Lorentz force, N or kN
   \item[$h_{k}$] specific enthalpy of $k$th species excluding  heat of formation, J/kg
   \item[$h_{k}^0$] heat of formation of $k$th species, J/kg
   \item[$H$] height of the computational domain, m
   \item[$I$] current flowing in the MHD system, A
   \item[$\vec{J}$] current density vector, $\mathrm{A/m^2}$
   \item[$J_{\rm emit}$] current density of thermionic emission, $\mathrm{A/m^2}$%?
   \item[$k_{\rm B}$] Boltzmann constant, J/K or erg/K
   {
   \item[$L_{\rm e-t}$] free electron -- translational energy relaxation distance, m
   \item[$L_{\rm v-t}$] vibrational -- translational energy relaxation distance, m}
   \item[$L_{\rm s}$] magnet separation distance, m
   \item[$L_{\rm m}$] magnet length along $x$, m
   \item[$L$] computational domain length along $x$
   {\item[$\ln \Lambda$] Coulomb logarithm}
   \item[$m_{\rm e}$] electron mass, kg
   \item[$m_{\rm i}$] mass of one ion particle, kg
   \item[M$_\infty$] freestream Mach number 
   \item[M] third bodies in reactions
   \item[$\ns$] number of species (including charged species)
   \item[$N$] total number density of the mixture, $\sum_k N_k$, $\mathrm{m^{-3}}$ or $\mathrm{cm^{-3}}$
   \item[$N_k$] species number density, $\mathrm{m^{-3}}$
   \item[$P$] sum of the partial pressures, Pa
   {
   \item[$P_{\rm dyn}$] dynamic pressure, $\frac{1}{2}\rho_\infty V_\infty^2$, Pa}
   \item[$P_\infty$] freestream pressure, Pa
   \item[$P_k$] species partial pressure, Pa
   {
   \item[$\overline{q_{\rm e}}$] electron thermal velocity, $\sqrt{8 k_{\rm B} T_{\rm e} / (\pi m_{\rm e})}$, m/s
   \item[$R$] specific gas constant, J/kgK
   \item[$r$] vector difference between the location where the induced magnetic field is measured and the location where it is produced, m}  
   \item[$s_k$] sign of the charge of species $k$ (either $+1$ for the positive species or $-1$ for the negative species)
   \item[$t$] time, s
   \item[$T$] bulk plasma temperature, K
   \item[$T_\infty$] freestream temperature, K
   \item[$T_w$] wall temperature, K
   \item[$\vec{V}$] velocity vector of the plasma bulk, m/s
   \item[$V_i$] velocity along $i$th dimension of the plasma bulk, m/s
   \item[$V_i^{k}$] velocity along $i$th dimension of the $k$th species, m/s
   {
   \item[$V_\infty$] freestream velocity, m/s}
   \item[$\mathrlap{-}V$] volume of the domain, m$^3$
   \item[$w_k$] mass fraction of $k$th species, $\rho_k/\rho$
   \item[$W$] work function, eV
   \item[$W_k$] mass production rate of species $k$ per unit volume due to chemical reactions, kg/$\mathrm{m^3}$/s
   \item[$x_i$] Cartesian coordinates, m
  \end{nomenclaturelist}
  
  \begin{nomenclaturelist}{Greek symbols}
  \item[$\beta_k^+$] {equal to 1 if species $k$ is a positive ion, 0 otherwise}
  \item[$\beta_k^{\rm n}$] equal to 1 if species $k$ is a neutral, 0 otherwise 
  \item[$\beta_k^{\rm c}$] equal to 1 if species $k$ is a charged species, 0 otherwise
  {
  \item[$\gamma$] specific heat ratio}
  {
  \item[$\gamma_{\rm e}$]  secondary electron emission coefficient}
  \item[$\Delta \phi$] voltage difference between the two magnet zones, V
  \item[$\epsilon_0$]   permittivity of free space, $\rm m^{-3} kg^{-1} s^4 A^2$
  \item[$\kappa$]       thermal conductivity of the mixture, W/mK
  \item[$\eta_{\rm MHD}$] MHD system efficiency defined as the ratio between the measured Lorentz force and the one of the ideal Lorentz force
  \item[$\eta_{\rm MHD,~theory}$] MHD system efficiency defined as the ratio between the Lorentz force obtained through a proposed theoretical expression and the one of the ideal Lorentz force
  \item[$\lambda_R$] material specific constant in Richardson's emission law
  {
  \item[$\mu_0$]        permeability of free space, $1.25663706\cdot10^{-6} ~\mathrm{m\cdot kg \cdot s^{-2} \cdot A^{-2}}$ }
  \item[$\mu_e$]        electron mobility, $\mathrm{m^2/Vs}$
  \item[$\mu_k$]        mobility of the $k$th species, $\mathrm{m^2/Vs}$
  \item[$\nu_k$]        mass diffusion coefficient for the neutral species, $\mathrm{m^2/s}$ or $\mathrm{cm^2/s}$
  \item[$\wtilde{\mu}^k$] tensor mobility of the $k$th species, $\mathrm{m^2/Vs}$
  \item[$\rho$]       mass density, kg/$\mathrm{m^3}$
  {
  \item[$\rho_\infty$]       freestream mass density, kg/$\mathrm{m^3}$}
  \item[$\rho_{\rm c}$]       net charge density, C/$\mathrm{m^3}$
  \item[$\rho_k$]       partial mass density of $k$th species, kg/$\mathrm{m^3}$
  \item[$\sigma$]       electrical conductivity, S/m
  \item[$\tau_{ji}$]  viscous stress in the direction of the $x_i$ coordinate and acting on surface perpendicular to $x_j$, $\mathrm{N/m^2}$
  {
  \item[$\tau_{\rm v-t}$]  vibrational-translational energy relaxation time, s}
  \item[$\chi$]       coordinate perpendicular to surface, m
  \item[$\phi$]       electric field potential, V
  \end{nomenclaturelist}
}

\begin{document}
\maketitle
\makenomenclature
\blfootnote{Presented as Paper AIAA-2024-1649 at the 2024 AIAA SciTech Forum, Orlando, FL, 8–12 January 2024}

\section{Introduction}

\dropword  Aerocapture presents an alternative to traditional propulsion techniques for spacecraft trajectory control \cite{jsr:1985:walberg}. Traditional propulsion methods involve the use of thrusters which can be problematic due to the additional propellant mass required to be carried on board the spacecraft. An aerocapture system offers a potential solution to these challenges by employing a maneuver that briefly enters the planet’s atmosphere to generate aerodynamic drag and/or lift forces. These aerocapture forces can be harnessed to alter the spacecraft trajectory or slow it down sufficiently to transition from atmospheric entry into orbit. In Ref.~\cite{jsr:2005:hall}, a cost-benefit analysis is conducted by comparing the initial mass to final mass ratio of a spacecraft during orbit insertion for both an aerocapture system and a propulsive system. The analysis concludes that an aerocapture system is significantly more cost-effective than alternative methods due to the reduced need for onboard fuel. However, subsequent studies (e.g., Refs. \cite{aiaa:2004:masciarelli, aiaa:2004:queen, aiaa:2015:lu, aiaa:2019:roelke, aiaa:2020:matz}) have revealed limited drag and lift forces, which in turn, have hindered sufficient deceleration for orbit insertion.

\begin{figure}[ht!]
     \centering
     \subfigure[Two-electrode MHD patch]{\includegraphics[width=0.4\textwidth]{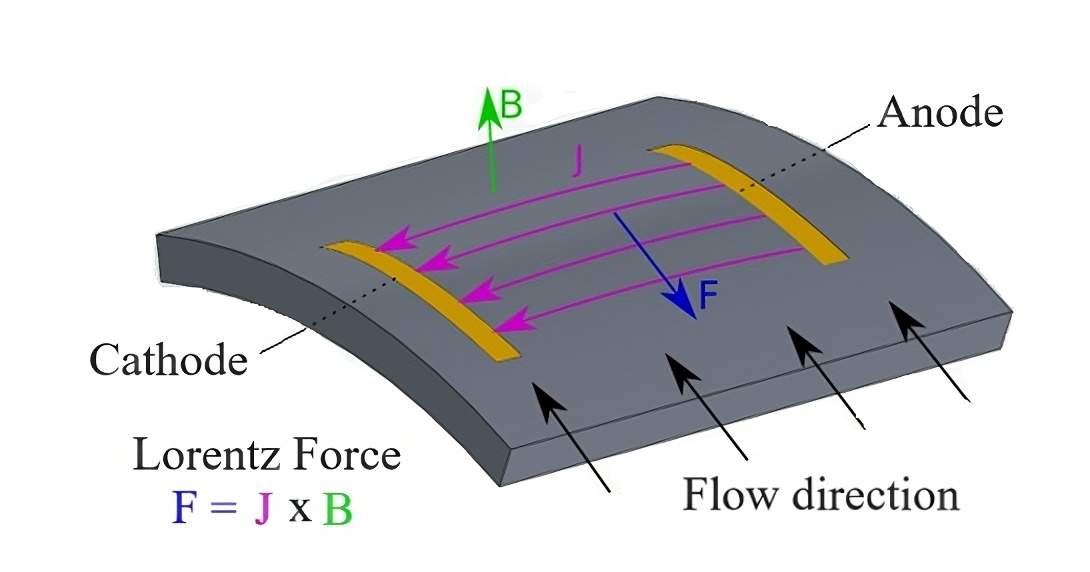}}
     \hfill
     \subfigure[Electrodeless MHD patch]{\includegraphics[width=0.34\textwidth]{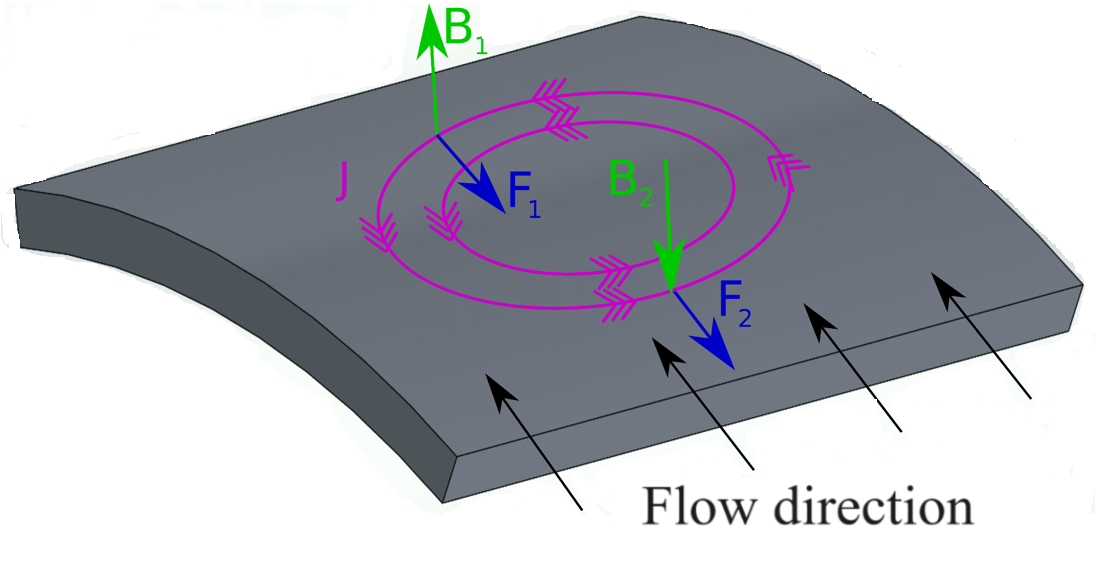}}
     \figurecaption{MHD patch configurations for aerocapture: (a) single-magnet two-electrode and (b) two-magnet electrodeless.}
     
     \label{fig:MHD_systems_schematic}
\end{figure}   

Recently, an improved approach to aerocapture has been proposed by Moses et al. \cite{aiaaconf:2022:moses}, which augments aerodynamic forces with Lorentz forces generated by a magnetohydrodynamics (MHD) patch positioned between the stagnation point and the aft of the capsule. Moses's MHD concept consists of utilizing a magnetic field that is perpendicular to and points outward from the aeroshell, coupled with two electrodes positioned in a manner that the current between them is perpendicular to both the flow velocity and the magnetic field vector (as depicted in Figure \ref{fig:MHD_systems_schematic}a). Given that the plasma flow possesses high electrical conductivity and velocity and interacts with a magnetic field oriented perpendicular to the flow, a flow of current ensues between the two electrodes via the Faraday electromotive force (EMF) principle. This leads to a strong Lorentz force  oriented perpendicular to both the current and magnetic field vectors and which can be harnessed for aerodynamic lift/drag manipulation.

Subsequent comprehensive fully-coupled numerical analyses of the Moses two-electrode MHD patch concept by Parent et al.\  revealed that a magnetic field strength of 0.1 Tesla or less could potentially generate a force comparable to that of aerodynamic forces \cite{jtht:2023:parent}. In the case of such low magnetic fields, the weight of the magnet is anticipated to be relatively minimal and not a major cause for concern. However, it was also observed that the plasma sheath that forms on the electrodes of a two-electrode MHD system could significantly impede the flow of current when thermionic emission is limited. In Refs.~\cite{aiaaconf:2022:parent,jtht:2023:parent}, it was demonstrated that the sheath diminishes the MHD force by 20 times or more when the electrode material shifts from thoriated tungsten (which exhibits high thermionic emission) to graphite (which exhibits low thermionic emission).

There are several drawbacks associated with utilizing electrodes in a MHD system for aerocapture. Firstly, the electrode short-circuit configuration leads to a concentration of high current and Joule heating near the edges of the electrodes, resulting in high temperatures in these regions. This necessitates the installation of additional cooling systems to manage these high temperatures. Moreover, the hot plasma flow generated during Earth entry can oxidize the tungsten electrodes due to the presence of oxygen in Earth’s atmosphere {\cite{jvsta:1998:bailey}}. This oxidation is problematic  as it could increase the work function of the tungsten and consequently lead to significant losses in thermionic emission. When thermionic emission is diminished, the current in the sheath shifts from being primarily carried by electrons (referred to as the inverse sheath regime) to being mostly carried by ions (referred to as the classical sheath regime). Since ions have a much lower mobility compared to electrons, this results in a substantial increase in resistance to current flow and consequently leads to significantly reduced MHD forces.

We here propose a novel solution to address these challenges by employing an electrodeless MHD system. As illustrated in Fig.~\ref{fig:MHD_systems_schematic}b, this system would incorporate two magnets oriented in opposite directions. This setup would cause the Faraday EMF to induce a current in a closed loop across the MHD patch, generating a large Lorentz force opposing the fluid flow. This new approach preserves the same advantages as Moses's two-electrode MHD aerocapture system over conventional aerocapture systems. These advantages include more efficient aerocapture than what can be achieved through standard aerodynamics alone. Moreover, the electrodeless MHD system is superior to the two-electrode MHD aerocapture system as it avoids issues such as tungsten electrode oxidation upon entry into Earth's atmosphere or the necessity for additional cooling systems to manage high current concentrations near the edges of electrodes.

In this paper, we will assess for the first time the performance of an electrodeless local MHD system compared to a two-electrode system. We will do so using fully-coupled 3D numerical simulations of a MHD patch located midway between the stagnation point and the aft of a planetary-entry capsule entering the Earth atmosphere at Mach 35. Here, a ``coupled simulation'' refers to the simulation of the electric field potential equation coupled with the Navier-Stokes equations for the bulk of the flow coupled with the drift-diffusion model for the charged species velocities as well as additional energy transport equations.   The results are obtained using the code CFDWARP {(Computational Fluid Dynamics, Waves, Reactions, Plasmas)} which has the unique capability to simulate efficiently the non-neutral sheaths near the electrodes in coupled form with the quasi-neutral bulk MHD flow \cite{jcp:2015:parent,aiaa:2016:parent}.

\section{Physical Model}

The physical model used herein thus consists of the drift-diffusion model for the charged species coupled to the Navier-Stokes equations for the neutral species. We here outline briefly the set of transport equations.

The conservation of mass for each charged species can be expressed as a function of the charged species velocity $\vec{V}^k$ which includes both drift and diffusion:
\begin{equation}
\frac{\partial}{\partial t} \rho_k + \sum_i \frac{\partial }{\partial x_i}\rho_{k} V_i^{k} = W_{k}  
\label{e:bparent:ddmass}
\end{equation}
where $\rho_k$ is the mass density of the $k$th species and where $W_k$ is the chemical source term consisting of the creation and destruction of the $k$th species due to chemical reactions.

For the drift-diffusion model the charged species velocity in the presence of a magnetic field becomes (see derivation in Ref.\ \cite{book:2022:parent}):
\begin{equation}
  V^{k}_i = V_i+\underbrace{\sum_j s_k \tilde{\mu}^k_{ij}  \left(\vec{E}  + \vec{V} \times \vec{B}\right)_j}_{\rm drift}
             - \underbrace{\sum_j  \frac{\tilde{\mu}^k_{ij}}{|C_k| N_k} \frac{\partial P_k}{\partial x_j}}_{\rm diffusion}
\label{eqn:ddmom}
\end{equation}
In the latter, $s_k$ is the species charge sign (either -1 for the negatively charged species or +1 otherwise), while $\vec{E}$, $\vec{B}$, and $\vec{V}$ are the electric  field, magnetic field, and bulk mixture velocity vectors, respectively. Also, the partial pressure $P_k$ is obtained from:
\begin{equation}\label{eq:idealgaslaw}
P_k= N_{\rm k} k_{\rm B}  T
\end{equation}
with $T$ the gas temperature (we here assume thermal equilibrium between all energy modes including vibrational energy and electron energy). Further, $C_k$ is the species charge ($+e$ for the  positive ions and $-e$ for the electrons). 
In the above,  the tensor mobility $\tilde{\mu}$ can be shown to be equal to \cite{book:2022:parent}:

\begin{align}
\tilde{\mu}^k  &\equiv \mu_k \left[\begin{array}{ccc} 
      1 & -s_k \mu_k  B_3  &  s_k \mu_k  B_2 \\
      s_k \mu_k  B_3 &  1 &  -s_k \mu_k  B_1 \\
      -s_k \mu_k  B_2 &  s_k \mu_k  B_1 & 1  \\
    \end{array} \right]^{-1}\nonumber\alb
&=\frac{\mu_k}{1+\mu_k^2 \vec{B}^2}\nonumber\alb  &\times\left[\!\!\begin{array}{ccc} 
      1+\mu_k^2  B_1^2 & \mu_k^2 B_1 B_2+s_k \mu_k  B_3  & \mu_k^2 B_1 B_3-s_k \mu_k  B_2   \\
      \mu_k^2 B_1 B_2-s_k\mu_k B_3 &  1+\mu_k^2 B_2^2 &  \mu_k^2 B_2 B_3+s_k\mu_k B_1 \\
      \mu_k^2 B_1 B_3 +s_k\mu_k B_2 & \mu_k^2  B_2 B_3-s_k\mu_k B_1  & 1+\mu_k^2 B_3^2  \\
    \end{array} \!\!\!\!\right]
\label{eqn:mutilde}
\end{align}
with $\mu$ the mobility.

In the above, the magnetic field is given a specified spatial distribution and is not altered as the iteration count progresses. {The magnetic field $\vec{B}$ thus corresponds to the sum of the applied magnetic field (from the magnets) and the induced magnetic field (from the plasma currents).} Further, because the magnetic field distribution does not vary in time, the induced electric field is negligible and we can use a potential equation based on Gauss's law to obtain the electric field as follows:
\begin{equation}
    \nabla^2 \phi = -\frac{\rho_{\rm c}}{\epsilon_0}
\end{equation}
with $\phi$, $\rho_{\rm c}$, and $\epsilon_0$   the electric field potential,  the net charge density,  and the permittivity of free space, respectively. From the potential we can find the electric field by taking the negative of its gradient (e.g. $\vec{E}=-\vec{\nabla} \phi$).

For the neutral species, a different mass conservation equation is used, as follows:
\begin{equation}
  \frac{\partial \rho_k}{\partial t} + \sum_{i=1}^3 \frac{\partial}{\partial x_i} \rho_k V_i  
- \sum_{i=1}^3 \frac{\partial}{\partial x_i} \left(\nu_k \frac{\partial w_k}{\partial x_i}\right)= W_k
\label{eqn:massneutral}
\end{equation}
where $w_k$ is the mass fraction of the $k$th species and $\nu_k$ is the diffusion coefficient. The momentum equation for the bulk of the plasma includes source terms to account for the electromagnetic body forces:
\begin{equation}
   \frac{\partial  }{\partial t}\rho V_i
  + \sum_{j=1}^3  \frac{\partial }{\partial x_j}\rho V_j V_i
=
-\frac{\partial P }{\partial x_i}  
+ \sum_{j=1}^3 \frac{\partial \tau_{ji}}{\partial x_j}
+ \rho_{\rm c} E_i + (\vec{J} \times \vec{B})_i
\label{eqn:mombulk}
\end{equation}
with $P$ being the sum of all partial pressures including the electron pressure and $\vec{J}$ being the current density equal to $\vec{J}\equiv \sum_k N_k \vec{V}^k C_k$. Also, $\tau_{ji}$ is the Navier stress tensor.

The temperature $T$ is determined from the total energy equation:
\begin{align}
& \frac{\partial }{\partial t}\rho e_{\rm t}
+ \sum_{j=1}^3  \frac{\partial }{\partial x_j} V_j \left(\rho  e_{\rm t} +  P \right)
- \sum_{j=1}^3  \frac{\partial }{\partial x_j} \left(
    \sum_{k=1}^{n_{\rm s}} \beta_k^{\rm n} \nu_k (h_k+h_k^\circ)\frac{\partial w_k}{\partial x_j} 
\right)
\nonumber \alb
&+ \sum_{j=1}^3 \sum_{k=1}^{n_{\rm s}}  \frac{\partial }{\partial x_j} \left(
  \beta_k^{\rm c} \rho_k (V^k_j-V_j) {(h_k+h_k^\circ)}
\right)
-\sum_{i=1}^{3}\frac{\partial }{\partial x_i}\left(\kappa \frac{\partial T}{\partial x_i} \right)
\alb
&=
 \sum_{i=1}^3 \sum_{j=1}^3  \frac{\partial }{\partial x_j} \tau_{ji} V_i
+ \vec{E}\cdot\vec{J} \nonumber
\label{eq:bulkenergy}
\end{align}
where  $\beta_k^{\rm n}$ is equal to 1 should species $k$ be neutral and to 0 otherwise and $\beta_k^{\rm c}$ is equal to 1 should species $k$ be charged and to 0 otherwise. Also, $\kappa$ is the sum of the thermal conductivities of all neutral and charged species. Further,  $e_{\rm t}$ corresponds to the total specific energy:
\begin{equation}
e_{\rm t} \equiv  \sum_{k=1}^{n_{\rm s}} w_k (e_k + h_k^\circ) +\frac{1}{2} \vec{V}^2
\end{equation}
In the latter, $h_k^\circ$ stands for the heat of formation while the sum $(h_k+h_k^\circ)$ represents the enthalpy of the $k$th species including calorically-imperfect effects as well as the heat of formation and is obtained from the McBride high temperature enthalpy polynomials \cite{nasa:2002:mcbride}. 

We are not including a separate transport equation for the vibrational energy and electron temperature in non-equilibrium in this preliminary study. {Such is an excellent approximation for the problems tackled herein because the vibrational-translational and electron-translational relaxation distances are at least one order of magnitude less than the length of the flow regions with significant MHD interaction. Indeed, for the problems under consideration, it can be easily shown that the vibrational-translational energy relaxation distance corresponds to (see derivation in Appendix A):
\begin{equation}
L_{\rm v-t} \approx \frac{1200 ~\textrm{Pa-m}}{P_{\rm dyn}}
\end{equation}
The flight dynamic pressure is here set to 50~kPa, hence resulting in a relaxation distance of approximately 2~cm. Such is more than 10 times less than the length of the MHD region, hence justifying our assumption of equilibrium between the vibrational and translational temperatures. A similar expression can be derived for the electron-translation energy relaxation distance (see derivation in Appendix B):
\begin{equation}
L_{\rm e-t} \approx \frac{0.2~\textrm{Pa-m} }{\frac{N_{\rm e}}{N} \cdot P_{\rm dyn}  }
\end{equation}
For the problems considered herein at Mach 35, the ionization fraction of the plasma $N_{\rm e}/N$ is 0.004 at the entrance of the MHD region and remains within this order of magnitude over the MHD region. Because the flight dynamic pressure considered here is of 50~kPa, the latter expression yields an electron-translational energy relaxation distance of approximately 1~mm. Such is several orders of magnitude less than the domain length, thus confirming the validity of our thermal equilibrium assumption. 
}

The transport coefficients (thermal conductivity, mobilities, viscosity, mass diffusion coefficients, etc) are taken from Ref.\ \cite{nasa:1990:gupta} but with some modifications as outlined in Ref.~\cite{jtht:2023:parent}. As well, we apply an additional correction {from Ref.~\cite{misc:1968:sinnott}} to take into account the effect of the electric field on the ion mobilities within the cathode sheath as outlined in Table \ref{tab:mobilities:Ecorrection}. 

As for the surface boundary conditions, we assume no surface catalysis for all neutral species. For the charged species, secondary electron emission and surface catalysis occurs when electrons and ions recombine at the surface. Both effects can be modelled through the following boundary conditions for electron and ion densities at the solid surfaces:
\begin{equation}
\frac{\partial }{\partial \chi} N_+ V^{+}_\chi = 0
{~~\rm and~~}
N_{-}=0
{~~\rm and~~}
N_{\rm e}=\frac{\gamma_{\rm e}}{\mu_{\rm e}} \sum_{k=1}^{n_{\rm s}} N_k \mu_k \beta_k^+
{~~\rm for~}
E_\chi<0
\end{equation}
\begin{equation}
N_{+}=0
{~~~~~\rm and~~~~~}
\frac{\partial }{\partial \chi} N_- V^{-}_\chi = 0
{~~~~~\rm and~~~~~}
\frac{\partial }{\partial \chi} N_{\rm e} V^{\rm e}_\chi= 0
{~~~~~\rm otherwise} 
\end{equation}
In the latter we use the subscripts ``+'', ``-'', and ``e'' to denote positive ions, negative ions, and electrons. Furthermore, $\chi$ denotes a coordinate perpendicular to the surface and oriented away from the surface toward the fluid, while the effective secondary electron emission {(SEE)} coefficient $\gamma_{\rm e}$ is varied between 1 and 100. {Although such values for the effective SEE coefficient are higher than the 0.1--0.3 often used for electrodes, we note that $\gamma_{\rm e}$ can sometimes have values in the range of 1--10 and even higher (see for instance experimental data in Ref.~\cite{psst:1999:phelps}). This is because incoming neutrals as well as incoming energetic electrons  \cite{psst:1999:phelps, misc:1956:kollath,misc:1958:dekker} can release electrons from the surface. Because  separate processes for secondary electron emission from neutral-electron or electron-electron collisions are not included here,  $\gamma_{\rm e}$ needs to account for all processes that result in secondary electron emission including ion-electron, electron-electron, and neutral-electron collisions at the surface. Consequently, our \emph{effective} electron yield per ion, $\gamma_{\rm e}$, is higher than it would be if it would only account for ion-electron surface collisions.}

We do not consider thermionic emission at the surface because thermionic emission is unlikely to be substantial even for materials with a low work function due to the oxidation of the electrodes by the air flow. { Indeed, a large loss of thermionic emission due to electrode oxidation has been observed experimentally  for  tungsten \cite{jvsta:1998:bailey} and for beryllium \cite{oom:1970:jerner} at low temperature. At higher air temperatures typical of planetary entry we expect even more oxidation to occur because of the higher concentrations of atomic oxygen and because the latter is more reactive than molecular oxygen.} Consequently, the electrons emitted from the surface are here restricted to secondary emission.

The chemical reactions are based on the 11-species air model outlined in Ref.~\cite{pof:2022:parent} but with modifications. The modifications include setting both the forward and backward temperatures to the gas temperature  and of computing the backward reaction rates using the equilibrium constant. The modified list of reactions and rates are given in Table \ref{tab:parent2023}. 
{It is emphasized that we do not assume chemical equilibrium. Rather, we use a Park-like non-equilibrium finite-rate chemical solver that is suited to planetary entry at high temperature through the use of the equilibrium constant to find the backward rates. The use of the equilibrium constant does not entail chemical equilibrium, but does lead to a much better prediction of the chemical composition of the plasma when the latter is close to chemical equilibrium as is the case here.}

In the cathode sheaths, the electron temperature reaches values orders of magnitude higher than the gas temperature. The important chemical reactions in the sheaths are the Townsend reactions (i.e., electron impact ionization) because they increase the electron and ion density and thus reduce the sheath resistivity to current flow significantly. Because we are not computing a separate transport equation for the electron temperature we can not make the Townsend reactions function of electron temperature. Thus, the Townsend reactions are expressed as a function of the reduced electric field as listed in Table \ref{tab:townsend}. Reactions 29-30 are taken from \cite{jcp:2014:parent} while Reactions 31-33 are obtained using BOLSIG+ \cite{psst:2005:hagelaar} with the Morgan cross-sections \cite{pcpp:1992:morgan}. When determining the various reactions using BOLSIG+ and the referenced cross sectional data, we made sure that the error does not exceed 30\% for the range of $E/N$ encountered in the problems herein.

\begin{table}[t]
  \center
  \begin{threeparttable}
    \tablecaption{Effect of electric field on electron and ion mobilities.\tnote{a}}
    \label{tab:mobilities:Ecorrection}
    \fontsizetable
    \begin{tabular}{lll}
    \toprule
    Species & Corrected Mobility, $\rm m^2\cdot V^{-1}\cdot s^{-1}$  & Reference\\
    \midrule
    N$_2^+$         & $\min\left( \mu_{\rm N_2^+},~~N^{-1}\cdot 2.03 \cdot 10^{12}\cdot \left(E^\star\right)^{-0.5} \right)$  & \cite{misc:1968:sinnott}\alb
    O$_2^+$         &  $\min \left( \mu_{\rm O_2^+},~~N^{-1}\cdot 3.61 \cdot 10^{12}\cdot\left(E^\star\right)^{-0.5}\right)$  & \cite{misc:1968:sinnott}\alb
    other~ions         & $\min\left( \mu_{\rm i},~~N^{-1}\cdot 0.55 \cdot \left(m_{\rm i} E^\star\right)^{-0.5}\right)$  & --\alb
    \bottomrule
    \end{tabular}
    \begin{tablenotes}
      \item[a] Notation and units:  $m_{\rm i}$ is the mass of one ion particule in kg; $N$ is the total number density of the plasma in 1/m$^3$; $E^\star$ is the reduced effective electric field  ($E^\star \equiv |\vec{E}|/N$) in units of V$\cdot$m$^2$; $\mu_k$ is the uncorrected mobility of species $k$ in $\rm m^2\cdot V^{-1}\cdot s^{-1}$.
    \end{tablenotes}
   \end{threeparttable}
\end{table}

\begin{table*}[!h]
\fontsizetable
\begin{center}
\begin{threeparttable}
\tablecaption{ 11-species high-temperature air plasma model.}
\begin{tabular*}{\textwidth}{c@{\extracolsep{\fill}}cccccc} 
\toprule
No. & Reaction\tnote{(b)} & Forward          & Backward    & $A$ & $n$ & $E$, cal/mole~\tnote{(a)}\\ 
~   & ~                   & Control. Temp.   & Control. Temp. &~ & ~ & ~ \\
\midrule
1 & $\rm N_2 + M_1 \rightleftarrows N + N + M_1$ & $T$ & $T$  & 3.0 $\cdot$ 10$^{22}$  & $-1.6$ & $113200 \, R$  \\

2 & $\rm N_2 + M_2 \rightleftarrows N + N + M_2$ & $T$ & $T$ & 7.0 $\cdot$ 10$^{21}$  & $-1.6$ & $113200 \, R$ \\

3 & $\rm N_2 + e^- \rightleftarrows N + N + e^-$ & $T$ & $T$ & 3.0 $\cdot$ 10$^{24}$  & $-1.6$ & $113200 \, R$  \\

4 & $\rm O_2 + M_1 \rightleftarrows O + O + M_1$ & $T$ & $T$ & 1.0 $\cdot$ 10$^{22}$  & $-1.5$ & $59500 \, R$ \\

5 & $\rm O_2 + M_2 \rightleftarrows O + O + M_2$ & $T$ & $T$ & 2.0 $\cdot$ 10$^{21}$  & $-1.5$ & $59500 \, R$ \\

6 & $\rm NO + M_3 \rightleftarrows N + O + M_3$ & $T$ & $T$ & 1.1 $\cdot$ 10$^{17}$  & 0.0 & $75500 \, R$ \\

7 & $\rm NO + M_4 \rightleftarrows N + O + M_4$ & $T$ & $T$ & 5.0 $\cdot$ 10$^{15}$  & 0.0 & $75500 \, R$ \\

8 & $\rm NO + O \rightleftarrows N + O_2 $ & $T$  & $T$ & 8.4 $\cdot$ 10$^{12}$  & 0.0 & $19400 \, R$ \\

9 & $\rm N_2 + O \rightleftarrows NO + N $ & $T$  & $T$ & 5.7 $\cdot$ 10$^{12}$  & $0.42$ & $42938 \, R$ \\

10 & $\rm N + O \rightleftarrows NO^+ + e^- $ & $T$ & $T$ & 5.3 $\cdot$ 10$^{12}$  & 0.0 & $32000 \, R$  \\

11 & $\rm O + O \rightleftarrows O_2^+ + e^- $ & $T$ & $T$ & 1.1 $\cdot$ 10$^{13}$  & 0 & $81200 \, R$ \\

12 & $\rm N + N \rightleftarrows N_2^+ + e^- $ & $T$ & $T$ & 2.0 $\cdot$ 10$^{13}$  & 0 & $67700 \, R$ \\

13 & $\rm NO^+ + O \rightleftarrows N^+ + O_2 $ & $T$  & $T$ & 1.0 $\cdot$ 10$^{12}$  & 0.5 & $77200 \, R$ \\

14 & $\rm N^+ + N_2 \rightleftarrows N_2^+ + N $ & $T$  & $T$ & 1.0 $\cdot$ 10$^{12}$  & 0.5 & $12200 \, R$ \\

15 & $\rm O_2^+ + N \rightleftarrows N^+ + O_2 $ & $T$ & $T$  & 8.7 $\cdot$ 10$^{13}$  & 0.14 & $28600 \, R$ \\

16 & $\rm O^+ + NO \rightleftarrows N^+ + O_2 $ & $T$ & $T$  & 1.4 $\cdot$ 10$^{5}$  & 1.90 & $26600 \, R$ \\

17 & $\rm O_2^+ + N_2 \rightleftarrows N_2^+ + O_2 $ & $T$ & $T$  & 9.9 $\cdot$ 10$^{12}$  & 0.00 & $40700 \, R$ \\

18 & $\rm O_2^+ + O \rightleftarrows O^+ + O_2 $ & $T$ & $T$  & 4.0 $\cdot$ 10$^{12}$  & $-0.09$ & $18000 \, R$ \\

19 & $\rm NO^+ + N \rightleftarrows O^+ + N_2 $ & $T$ & $T$  & 3.4 $\cdot$ 10$^{13}$  & $-1.08$ & $12800 \, R$ \\

20 & $\rm NO^+ + O_2 \rightleftarrows O_2^+ + NO $ & $T$ & $T$  & 2.4 $\cdot$ 10$^{13}$  & 0.41 & $32600 \, R$ \\

21 & $\rm NO^+ + O \rightleftarrows O_2^+ + N $ & $T$ & $T$  & 7.2 $\cdot$ 10$^{12}$  & 0.29 & $48600 \, R$ \\

22 & $\rm O^+ + N_2 \rightleftarrows N_2^+ + O $ & $T$ & $T$ & 9.1 $\cdot$ 10$^{11}$  & 0.36 & $22800 \, R$ \\

23 & $\rm NO^+ + N \rightleftarrows N_2^+ + O $  & $T$ & $T$ & 7.2 $\cdot$ 10$^{13}$  & 0.00 & $35500 \, R$  \\

24 & $\rm O + e^- \rightleftarrows O^+ + e^- + e^- $ & $T$  & $T$ & 6.37 $\cdot$ 10$^{16}$  & $0.0029$ & $477190 \, R$  \\

25 & $\rm N + e^- \rightleftarrows N^+ + e^- + e^- $ & $T$ & $T$ & 1.06 $\cdot$ 10$^{18}$  & $-0.2072$ & $629700 \, R$ \\

26 & $\rm O_2 + e^- \rightleftarrows O_2^+ + e^- + e^-$ &$T$ &$T$ & 2.33 $\cdot$ 10$^{16}$ & $0.1166$ & $567360\, R$ \\

27 & $\rm N_2 + e^- \rightleftarrows N_2^+ + e^- + e^-$ &$T$ &$T$ & 1.58 $\cdot$ 10$^{16}$ & $0.1420$ & $536330\, R$ \\

28 & $\rm NO + e^- \rightleftarrows NO^+ + e^- + e^-$ &$T$ &$T$ & 5.63 $\cdot$ 10$^{18}$ & $-0.2607$ & $686030\, R$ \\

\bottomrule
\end{tabular*}
\begin{tablenotes}
\item[{a}] The universal gas constant $R$ must be set to 1.9872	cal/K$\cdot$mol. $A$ has units of $\textrm{cm}^3\cdot(\textrm{mole}\cdot \textrm{s})^{-1}\cdot \textrm{K}^{-n}$. $E$ has units of cal/mole. The rate is given by $A T^n \exp(-E/RT).$
\item[{b}] $\rm M_1=N,~O,~N^+,~O^+$; $\rm M_2=N_2,~O_2,~NO,~N_2^+,~O_2^+,~NO^+$; $\rm M_3= N,~O,~NO,~N^+,~O^+$; $\rm M_4=N_2,O_2,N_2^+,~O_2^+,~NO^+$.
\end{tablenotes}
\label{tab:parent2023}
\end{threeparttable}
\end{center}
\end{table*}

\begin{table*}[t]
  \center\fontsizetable
  \begin{threeparttable}
    \tablecaption{Additional Townsend ionization and recombination rates for N$_2$, O$_2$, N, O, and NO.\tnote{a}}
    \label{tab:townsend}
    \fontsizetable
    \begin{tabular*}{\textwidth}{l@{\extracolsep{\fill}}lll}
    \toprule
    No.&Reaction & Rate Coefficient  & Refs. \\
    \midrule
    29  & $\rm e^- + N_2   \rightarrow N_2^+ + e^- + e^-$  
       &  ${\rm exp}(-0.0105809\cdot {\rm ln}^2 E^\star - 2.40411\cdot 10^{-75} \cdot {\rm ln}^{46}E^\star)$~cm$^3$/s
       & \cite{jcp:2014:parent} \\
    30  & $\rm e^- + O_2   \rightarrow O_2^+ + e^- + e^-$  
       &  ${\rm exp}(-0.0102785\cdot {\rm ln}^2 E^\star - 2.42260\cdot 10^{-75} \cdot {\rm ln}^{46}E^\star)$~cm$^3$/s
       & \cite{jcp:2014:parent} \\
    31 & $\rm e^- + N   \rightarrow N^+ + e^- + e^-$  
       &  ${\rm exp}(- 9.3740 \cdot 10^{-3}\cdot {\rm ln}^2 E^\star - 3.3250\cdot 10^{-23} \cdot {\rm ln}^{14}E^\star)$~cm$^3$/s
       & \cite{psst:2005:hagelaar,pcpp:1992:morgan} \\
    32  & $\rm e^- + O   \rightarrow O^+ + e^- + e^-$  
       &  ${\rm exp}(- 1.0729 \cdot 10^{-2}\cdot {\rm ln}^2 E^\star + 1.6762\cdot 10^{-87} \cdot {\rm ln}^{53}E^\star)$~cm$^3$/s
       & \cite{psst:2005:hagelaar,pcpp:1992:morgan} \\
    33  & $\rm e^- + NO   \rightarrow NO^+ + e^- + e^-$  
       &  ${\rm exp}(-5.9890\cdot 10^{-6} \cdot {\rm ln}^4 E^\star + 2.5988\cdot 10^{-84} \cdot {\rm ln}^{51}E^\star)$~cm$^3$/s
       & \cite{psst:2005:hagelaar,pcpp:1992:morgan} \\
    \bottomrule
    \end{tabular*}
\begin{tablenotes}
\item[{a}] Notation and units: ${\cal A}$ is Avogadro's number set to $6.02214 \times 10^{23}$ mol$^{-1}$; $E^\star$ is the reduced effective electric field ($E^\star\equiv|\vec{E}|/N$) in units of V$\cdot$m$^2$.

\end{tablenotes}
   \end{threeparttable}
\end{table*}

\section{Numerical Methods}

To relieve the stiffness of the non-neutral plasma flow equations outlined in the previous section, we employ the recast of the governing equations outlined in Refs.~\cite{book:2022:parent,jcp:2015:parent}. Then, we discretize the recast set of equations using the Roe scheme turned second-order accurate through the Van-Leer TVD limiter. We further use the eigenvalue conditioning based on the Peclet number outlined in Ref.~\cite{aiaa:2017:parent}. The Peclet-number-based eigenvalue conditioning is advantaged over others by being both capable  to prevent carbuncles and not affecting the resolution of the viscous layers. The discretized fluid flow transport equations are converged to steady-state using a DDADI block-implicit method while the electric field potential equation is converged using both a Successive Over Relaxation (SOR) algorithm and an Iterative Modified Approximate Factorization (IMAF) algorithm \cite{cf:2001:maccormack}. 

\section{Problem Setup}

\begin{figure}[!ht]
     \centering
     \includegraphics[width=0.33\textwidth]{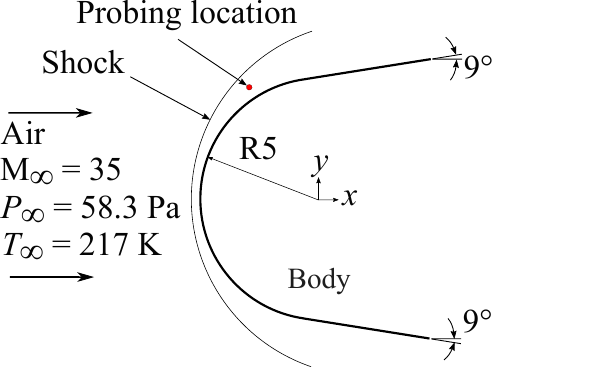}
     \figurecaption{Problem setup for the Mach 35 Earth entry case; dimensions in meters.}
     \label{fig:problem_setup}
\end{figure}

\begin{figure}[ht!]
     \centering
     \subfigure[]{\includegraphics[width=0.35\textwidth]{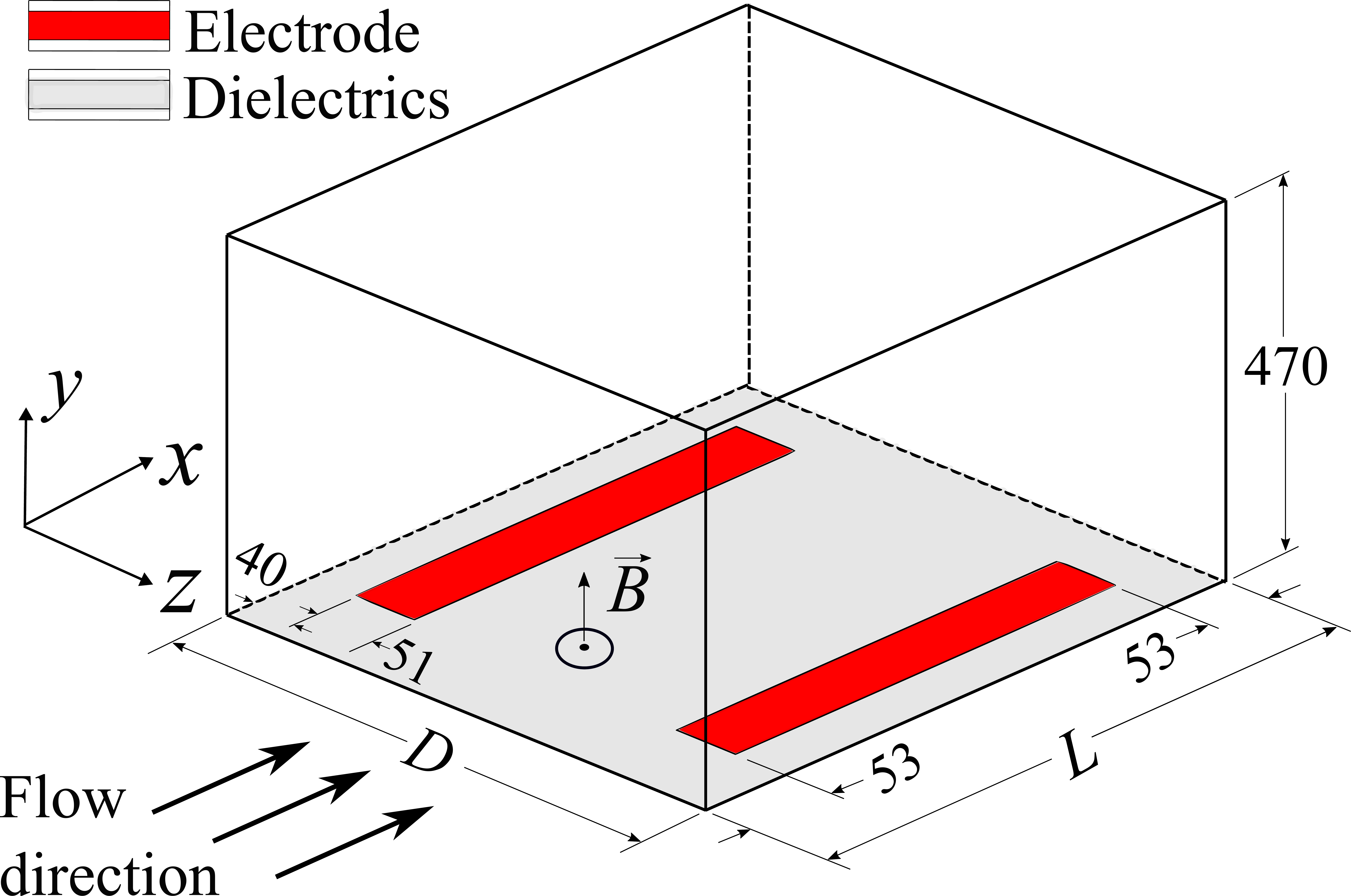}}
     \subfigure[]{\includegraphics[width=0.35\textwidth]{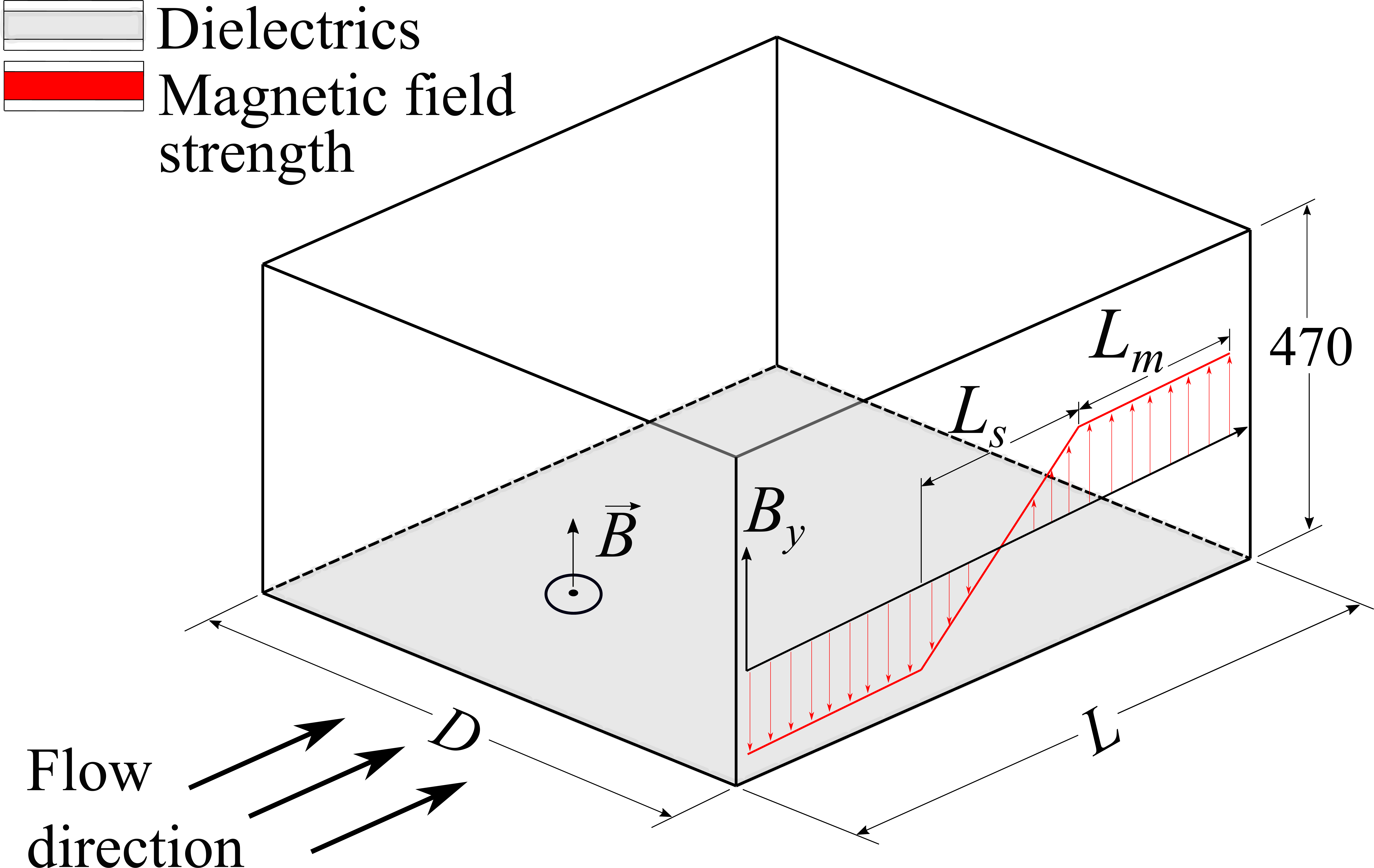}}
     \figurecaption{Problem setup for (a) the two-electrode configuration and (b) the two-magnet electrodeless configuration.}
     \label{fig:setupMHDpatch}
\end{figure}

First we obtain a steady-state solution of the 2D axisymmetric flow around a capsule entering the Earth atmosphere at Mach 35 with a geometry and freestream conditions as depicted in Fig.\ \ref{fig:problem_setup}. The inflow properties  for the two-electrode MHD patch shown in Fig.\ \ref{fig:setupMHDpatch}a are obtained by probing the flow speed, temperature, pressure and species concentrations at the specific location marked on the schematic. This results in an inflow static pressure of 29.4 kPa, an inflow static temperature of 7925 K and an inflow speed of 5090 m/s.  Non-slip wall boundary conditions are imposed on the bottom boundary and the wall temperature is fixed to 1500 K. Symmetry conditions are imposed on the front, back and top boundaries, while a supersonic outflow condition is imposed on the right boundary. The two electrodes in the baseline configuration are 400~mm in length and 51~mm in width, separated 40~mm from the sides of the domain of depth $D$, aligned to the flow direction and placed on a dielectric surface. A constant magnetic field strength $B_y$ is imposed on this domain and is set to 0.1~T. By varying the domain length $L$ and the depth $D$ while keeping all other parameters constant, the electrode spacing and length are altered. Unless otherwise indicated the domain length and depth are both fixed to 0.5~m. 

The electrodeless MHD patch uses the same domain and inflow conditions as the two-electrode system but without any electrodes present (see Fig.\ \ref{fig:setupMHDpatch}b). The bottom surface is thus a dielectric wall. In order for the current loop to create a force opposing the flow, the magnetic field vector must point in opposing directions over two portions of the domain. First, the magnetic field vector points in the negative $y$ direction and transitions to a positive component over a distance $L_{\rm s}$. The magnetic field strength varies linearly over the length $L_{\rm s}$ as follows (see also Fig.\ \ref{fig:setupMHDpatch}b):
\begin{equation}
B_y(x) = B\left[1-{\rm min}\left(2,{\rm max}\left[0,1-\dfrac{2}{L_{\rm s}}\left(x-\dfrac{L}{2}\right)\right]\right)\right]
\end{equation}
where $B_y(x)$ is the local value of the $y$ component of the magnetic field and $B$ is the user specified maximum magnetic field strength which is here set for all cases to 0.1~T. Along the $x$ and $z$ dimensions, the magnetic field vector is given a zero value. 

\section{Numerical Error}

\begin{figure}[!b]
     \centering
     \subfigure[]{\includegraphics[width=0.33\textwidth]{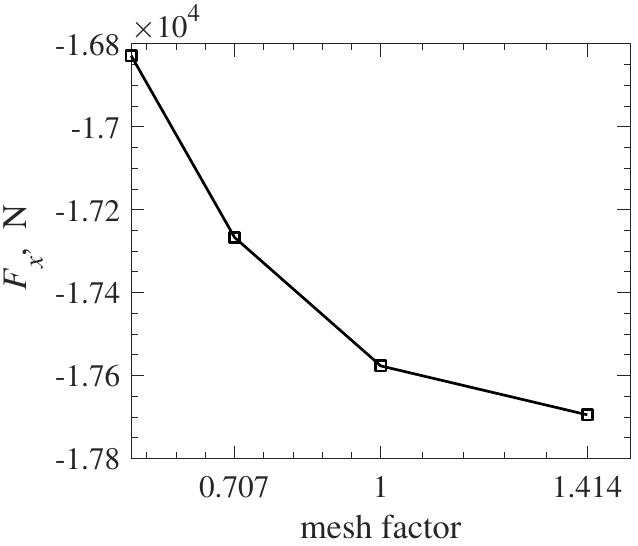}}
     \subfigure[]{\includegraphics[width=0.33\textwidth]{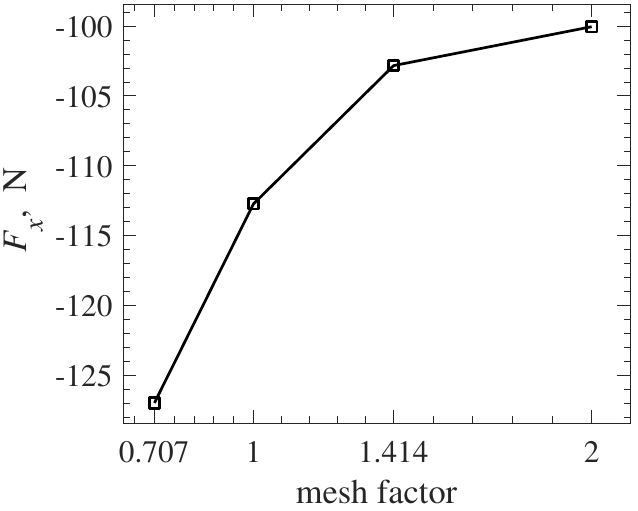}}
     \figurecaption{Effect of grid size on net Lorentz force for the  (a) the electrodeless case with $D=1.5$ m and $L_{\rm s}=0.5$ m and (b) two-electrode case with $D=0.5$ m and $\gamma_e = 10$.}
     \label{fig:gridconv}
\end{figure}

\begin{figure}[ht!]
     \centering
     {\includegraphics[width=0.33\textwidth]{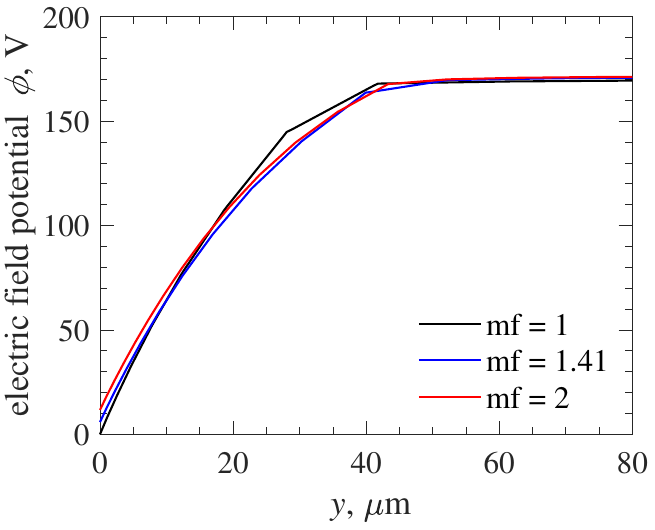}}
     \figurecaption{Effect of grid size on the electric potential profiles extracted from the cathode surface for the two-electrode case with  $\gamma_e=10$ and $D=0.5$~m.}
     \label{fig:gridconv_sheath}
\end{figure}

Several grid convergence studies are performed to obtain an estimate of the numerical error present in our results, both for the two-electrode configuration and for the electrodeless configuration. 

The first grid convergence study is focused on the electrodeless two-magnet configuration with the domain depth $D=1.5$~m and the magnet separation distance $L_{\rm s}=0.5$~m. The baseline mesh (with a mesh factor of 1) is composed of $79 \times 60 \times 150$ nodes. Coarser or finer meshes are created by reducing or increasing the mesh factor. Thus, for a mesh factor of 0.5, the mesh is composed of $40 \times 30 \times 75$ nodes while for a mesh factor of 1.41, it is composed of $111 \times 85 \times 212$ nodes. We assess the impact of the mesh size on the net force created by the MHD device in Fig.~\ref{fig:gridconv}a. Clearly, it can be observed that the net force converges towards a root as expected for a second-order accurate stencil. That is, as the mesh factor is increased two-fold, the error decreases four-fold. Using the Grid Convergence Index (GCI) \cite{aiaa:1998:roache}, and noting that the solution is in the asymptotic range of convergence, the error on the net Lorentz force obtained using the baseline mesh is calculated to be of 1\%.    To keep the numerical error this low for other configurations (e.g., a different domain depth or a different spacing between the magnets), we vary the mesh size such that the grid density remains the same as for the baseline mesh. Thus, when the domain depth is increased from 0.5~m to 1.5~m, the number of grid lines along $z$ is increased three-fold.  

Another grid convergence study is now performed for the two-electrode configuration (Fig.~\ref{fig:gridconv}b). The main difficulty in simulating this problem is to capture accurately the voltage drop within the non-neutral sheath near the cathode. Indeed, because the sheath is in the standard regime (i.e., it has a higher ion density than electron density and the current is mostly ionic rather than electronic), it leads to a high resistance to the current flow and thus to a large voltage drop that can be close to the Faraday electromotive force, effectively reducing considerably the current flow and, thus, the Lorentz forces. Independently of the SEE coefficient, the mesh is designed such that there is clustering at the electrodes with the near wall spacing fixed to 1 micrometer and grows exponentially as the distance from the wall increases. Such a grid spacing strategy is found to yield a very good resolution of the sheath resistance to current flow. Indeed, as can be seen from the electric field potential extracted through the cathode sheath shown in Fig.~\ref{fig:gridconv_sheath}, very little difference can be observed between the three different meshes used. This excellent performance of our CFDWARP plasma solver to resolve the sheath potential drop even on coarse meshes explains why we are able to obtain a small numerical error on the net Lorentz force of less than 1\% when the mesh factor is 2 (see Fig.\ \ref{fig:gridconv}b).   
%This explains why..   

\section{Results and Discussion}

In this section, a detailed comparison between the electrodeless and the two-electrode MHD patches is presented. We start by investigating the effect of the secondary electron emission on the efficiency of the two-electrode MHD patch. This is followed by a parametric study of the effect of MHD domain depth on the performance of both the two-electrode and the electrodeless configurations. Then,  we assess how sensitive the performance of the electrodeless system is to the spacing between the two magnets.

\subsection{Effect of Secondary Electron Emission}

\begin{table}[t]
  \center\fontsizetable
  \begin{threeparttable}
    \tablecaption{Effect of secondary electron emission coefficient on the MHD efficiency.\tnote{(a)}}
    \label{tab:parametric_gammae}
    \fontsizetable
\begin{tabular*}{0.49\textwidth}{c@{\extracolsep{\fill}}ccc}
\toprule
\multirow{1}*{$\gamma_{\rm e}$}& $I$, ${\rm A}$   &  $F_x$, $\rm N$  &$\eta_{\rm MHD} $ \\
\midrule
 $1$      & $202$ &   $-6$ & 0.001  \\
$10$      &$2820$ &    $-120$ & 0.02   \\
$100$     & $34000$ &  $-1340$ & 0.23   \\
\bottomrule
\end{tabular*}
\begin{tablenotes}
\item[{a}] The domain depth is set to $D=0.5$~m. 
\end{tablenotes}
   \end{threeparttable}
\end{table}

The performance of MHD systems utilizing electrodes may be significantly impacted by the non-neutral plasma sheaths present near the electrodes. This is mostly due to the non-neutral sheath that develops over the cathode, characterized by an electron density much lower than the ion density. This  results in ions carrying most of the current within the sheaths. Given that ion mobilities are orders of magnitude lower than electron mobility, this creates a substantial resistance to current flow, thereby diminishing MHD performance. For example, in a recent study it was found that changing the electrode material from graphite to thoriated tungsten resulted in a tenfold increase in MHD forces \cite{jtht:2023:parent}. This increase was attributed to thoriated tungsten's higher thermionic emission coefficient compared to graphite, which shifted the sheath regime from classical (more ions than electrons) to the inverse regime (more electrons than ions) by populating the sheath with numerous electrons. Consequently, thoriated tungsten significantly reduced the sheath's resistance to current flow by orders of magnitude and lead to much higher Lorentz forces.

In this paper, we use electrode boundary conditions that include secondary electron emission (SEE) but not thermionic emission. Because the secondary electron emission coefficient $\gamma_{\rm e}$ is not well known for the plasma under consideration, we here investigate its impact on the MHD forces. 

\begin{figure}[ht!]
     \centering
     \subfigure[]{\includegraphics[width=0.37\textwidth]{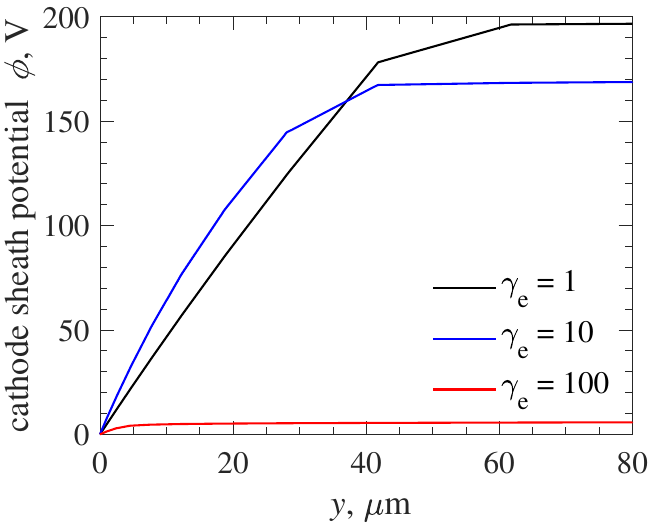}}
     % \subfigure[]{\includegraphics[width=0.32\textwidth]{figures_AIAAJ_article/gammae_rhoc_4.pdf}}
     \subfigure[]{\includegraphics[width=0.38\textwidth]{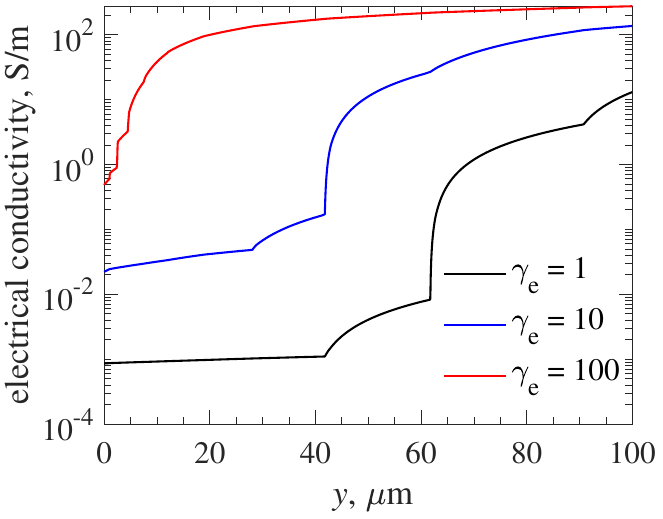}}     
     \figurecaption{Effect of secondary electron emission coefficient on (a) cathode sheath voltage and (b) electrical conductivity  with $D=0.5$~m.}
     \label{fig:gammae_rhoc_phi}
\end{figure}

\begin{figure*}[ht!]
     \centering
     \subfigure[$\gamma_{\rm e} = 1$]{\includegraphics[width=0.29\textwidth]{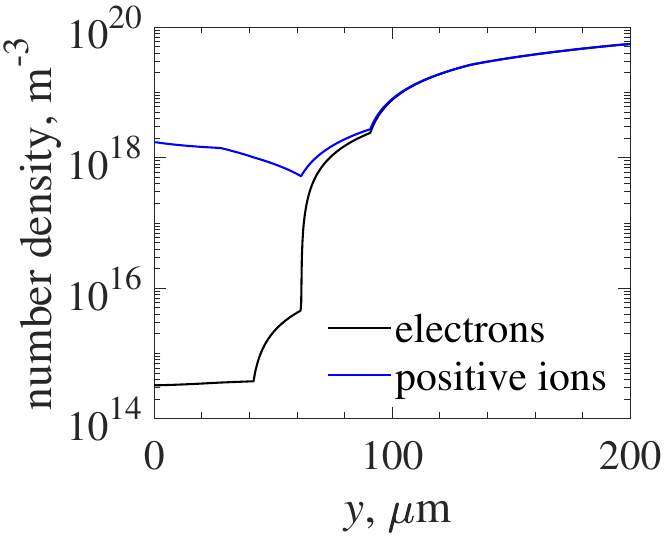}}
     \subfigure[$\gamma_{\rm e} = 10$]{\includegraphics[width=0.29\textwidth]{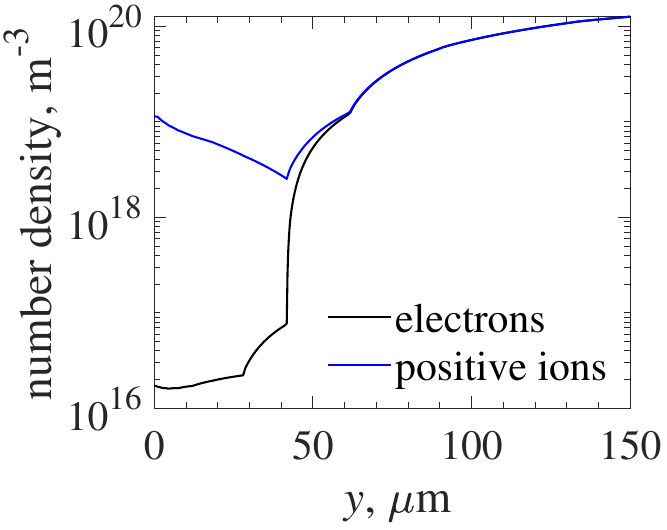}}
     \subfigure[$\gamma_{\rm e} = 100$]{\includegraphics[width=0.29\textwidth]{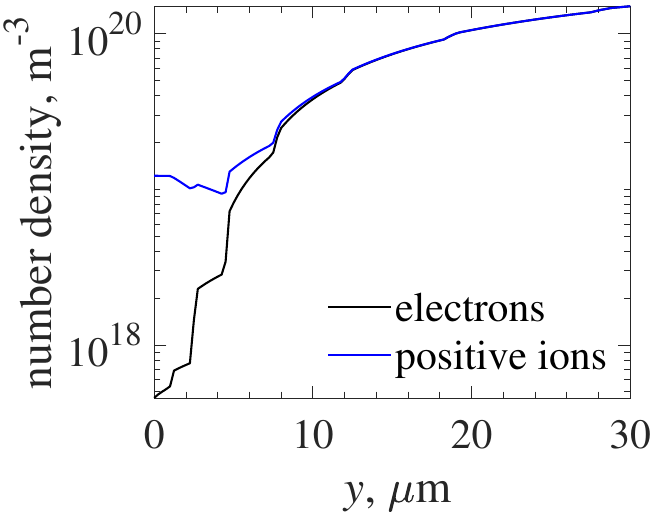}}     
     \figurecaption{Effect of secondary electron emission coefficient on electron and ion number density in the cathode sheath with $D=0.5$~m.}
     \label{fig:gammae_ne_ni}
\end{figure*}

As evident from Table \ref{tab:parametric_gammae}, a $\gamma_{\rm e}$ variation from 1 to 10 results in a more than tenfold increase in MHD forces. This is primarily due to an increase in the sheath's resistance to current flow when the secondary electron emission coefficient is reduced. This is corroborated by Fig.\ \ref{fig:gammae_rhoc_phi}b, which illustrates that the average electrical conductivity in the sheath is approximately ten times lower when the SEE coefficient decreases from 10 to 1 due to a larger sheath thickess (see Fig.\ \ref{fig:gammae_ne_ni}). However, it is not entirely clear why the voltage drop through the sheath does not exhibit the same trend. Intuitively, one might expect the voltage drop to increase more or less in proportion to the sheath's resistance. However, this is not observed. As shown in Fig.\ \ref{fig:gammae_rhoc_phi}a, the voltage drop in the sheath only varies from 200 volts to 170 volts when the secondary electron emission increases tenfold from 1 to 10. Why is there a relatively small difference in voltage despite a significant change in the sheath's resistivity?

The voltage difference across the cathode sheath changes by only 15\% despite the sheath's resistivity decreasing by one order of magnitude when the SEE coefficient changes from 1 to 10. This is because the current density also varies by nearly one order of magnitude. Ultimately, the voltage across the sheath cannot exceed the Faraday EMF, which corresponds to 
\begin{equation}
 \mathcal{E} = V_x L_z B_y
 \label{eqn:FaradayEMF}
\end{equation}
with $\mathcal{E}$ the Faraday EMF (in Volts), $V_x$ the streamwise plasma bulk velocity, $L_z$ a characteristic distance along $z$ through which the current travels, and $B_y$ the component of the magnetic field perpendicular to the surface. For the MHD patch problems studied herein, the flow velocity is approximately 5000 m/s, and the distance the current travels from one electrode edge to the other, $L_z$, is roughly 0.4 m when the domain depth is set to 0.5 m. Given a magnetic field of 0.1 T, this calculates to a Faraday EMF of 200 V. It is noted that while the current is driven by the Faraday EMF, it is restrained by the voltage difference across the sheaths. Assuming that no voltage is lost within the anode sheath (a very good assumption due to its conductivity being as high as in the nearby plasma), the degree to which the Faraday EMF is diminished is related to the voltage difference across the cathode sheath. Indeed, if the cathode sheath were to possess infinite resistivity, no current would flow, and the voltage drop across the cathode sheath would equal the Faraday EMF. For the problem addressed herein, the maximum voltage difference that the cathode sheath might undergo is 200 V. Thus, one would anticipate a voltage drop of nearly 200 V when the sheath poses significant resistance to the current flow. This is precisely what is observed in Fig. \ref{fig:gammae_rhoc_phi}a when the SEE coefficient is low. As the SEE coefficient is increased, the sheath-induced voltage impedes the Faraday EMF less. This allows more current to flow from one electrode to the other hence resulting in a higher Lorentz force. This pattern is discernible in Table \ref{tab:parametric_gammae}, where the Lorentz force increases almost tenfold when the SEE coefficient is diminished by the same amount.

Due to the high temperature of the electrodes (1500 K), it is possible for the electrode material to emit electrons through thermionic emission, thereby reducing the sheath resistance. However, we do not incorporate this effect in our paper for two reasons: firstly, it is improbable to be significant due to the oxidation of the electrodes by atomic oxygen, and secondly, even if oxidation were not to alter the thermionic capabilities of the electrode, such an effect is unlikely to substantially impact the results in this case. To assess the potential impact of thermionic emission in this scenario, Richardson's thermionic emission law \cite{sse:1965:crowell} can be used as follows:
\begin{equation}
    J_{\rm emit}= \lambda_R A_0 T_{\rm w}^2 \exp\left(-\dfrac{W}{k_{\rm B} T_{\rm w}}\right) 
\end{equation}
where $T_{\rm w}$ is the wall temperature in Kelvin, $W$ is the work function in eV, $k_{\rm B}$ is the Boltzmann constant, $\lambda_R$ is a material specific constant here set to 0.5, and $A_0$ is a constant set to $1.202\cdot10^6$~A/m$^2$K$^2$. With an electrode temperature of 1500 K and a work function of 2.4 (for thoriated tungsten), Richardson's law would predict an emitted current of 10,000 Amps per square meter. However, this is unlikely to have a significant impact on the results, particularly for a secondary electron emission (SEE) coefficient of 10 or higher. This is due to the fact that the average current density on the electrodes in these cases is much greater (over 1 million Amps per square meter for a SEE coefficient of 100), rendering the contribution from thermionic emission relatively insignificant.

It could be argued that the SEE coefficient is typically set to 0.1 for many problems, rather than 1 or higher as in our case. However, we would like to emphasize two points: firstly, the effective SEE coefficient is not well established for the gas composition considered in our study, and secondly, the limited experimental data on effective SEE for other gases suggest that it varies significantly with the electric field. For instance, for argon, the effective SEE coefficient has been observed to vary between 0.01 and 10 when the reduced electric field increased from 1000 Townsend (Td) to 1 million Td. This trend was observed across various electrode materials such as copper, brass, nickel, and steel \cite{psst:1999:phelps}. In our problem, the maximum reduced electric field near the cathode typically exceeds 10,000 Td, which justifies our chosen range of $\gamma_e$ in excess of 1.

\subsection{Effect of MHD Domain Depth}

\begin{table}[t]
  \center\fontsizetable
  \begin{threeparttable}
    \tablecaption{Effect of domain depth on the MHD efficiency.}
    \label{tab:F_vs_domain_depth}
    \fontsizetable
\begin{tabular*}{0.46\textwidth}{@{}c@{\extracolsep{\fill}}cccccc@{}}
\toprule
~&\multicolumn{2}{c}{Two-Electrode\tnote{(a)}}& \multicolumn{1}{c}{}  &\multicolumn{2}{c}{electrodeless\tnote{(b)}} \\
\cmidrule{2-3}\cmidrule{5-6}
$D,~\rm m$&$F_{x}$, kN&$\eta_{\rm MHD}$& & $F_{x}$, kN& $\eta_{\rm MHD}$   \\
\midrule
0.50&   -1.34  & 0.23 &   &  $ -3.22$   &  0.54\\
1.50&   - 12.77& 0.72 &  &  $-16.55$ &  0.93  \\
\bottomrule
\end{tabular*}
\begin{tablenotes}
\item[{a}] The secondary electron emission coefficient is fixed to $100$.
\item[{b}] The magnet separation distance is fixed to $L_{\rm s}=0.01$~m.
\end{tablenotes}
   \end{threeparttable}
\end{table}

\begin{figure}[ht!]
     \centering
     \subfigure[]{\includegraphics[width=0.35\textwidth]{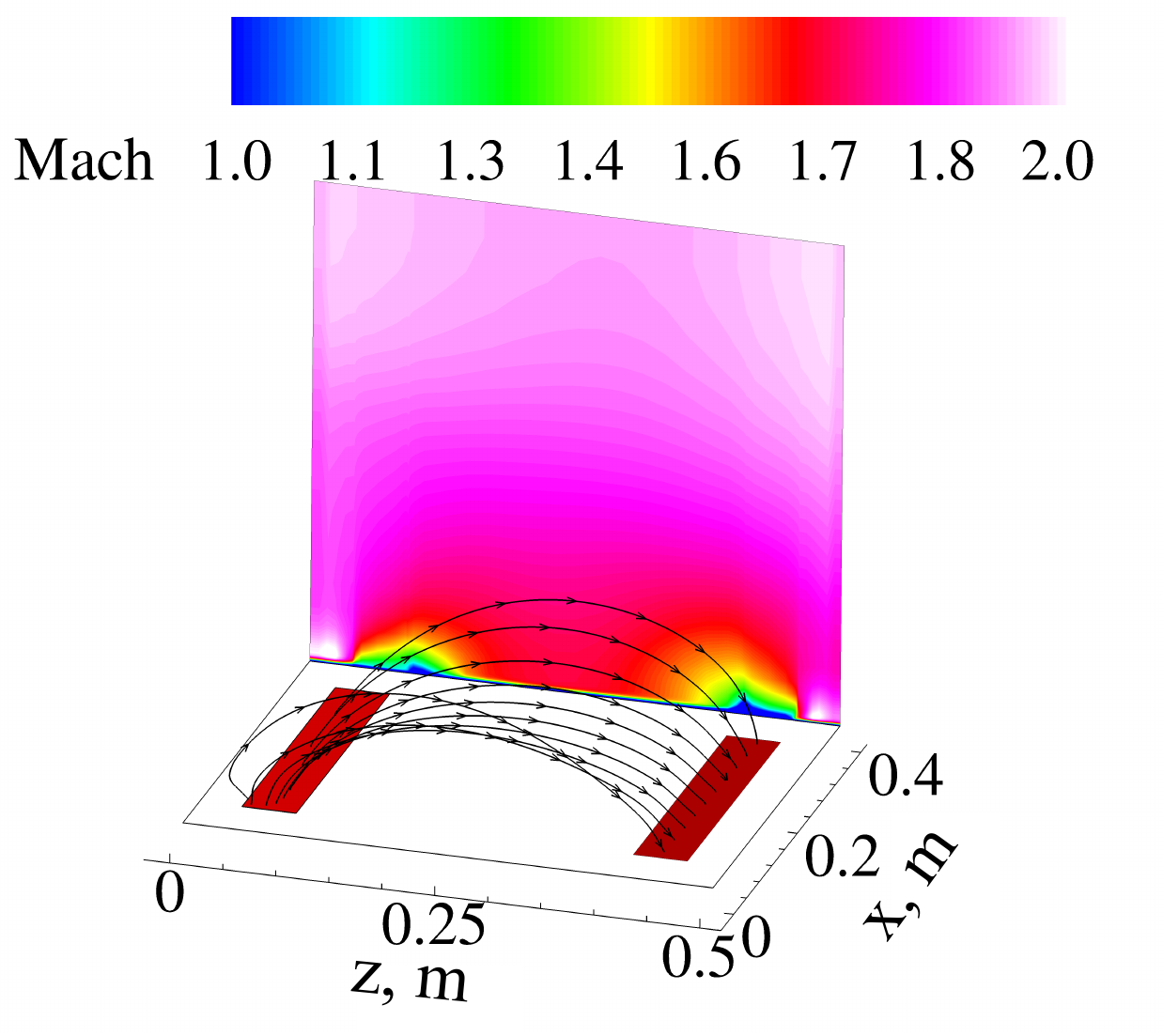}}
     \subfigure[]{\includegraphics[width=0.35\textwidth]{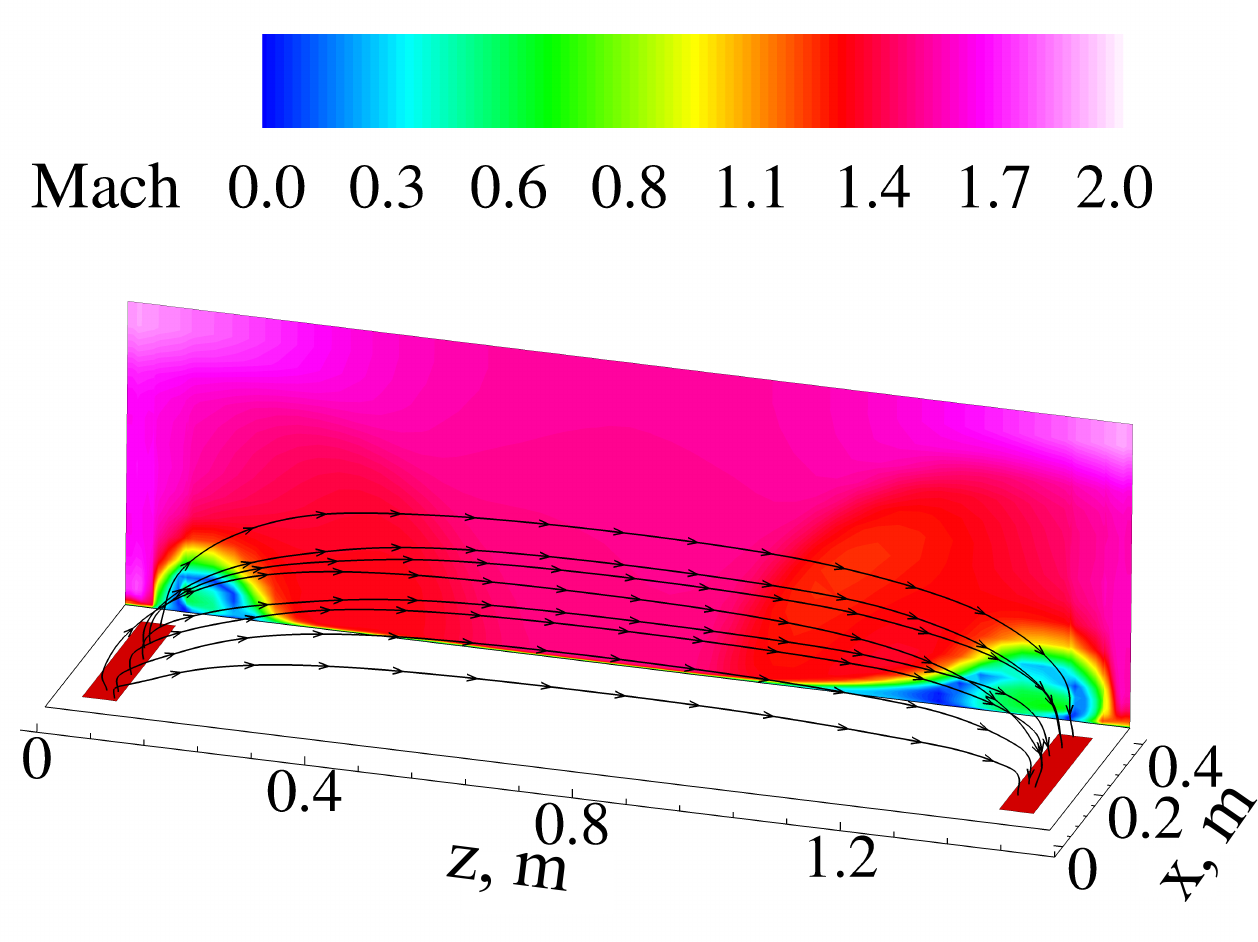}}
     \figurecaption{Current density streamlines and Mach number contours for the two-electrode configuration with  $\gamma_e=100$ for (a) $D=0.5$ m, and  (b) $D=1.5$~m.}
     \label{fig:mach_current_twoelectrode}
\end{figure}

\begin{figure}[ht!]
     \centering
     \subfigure[]{\includegraphics[width=0.33\textwidth]{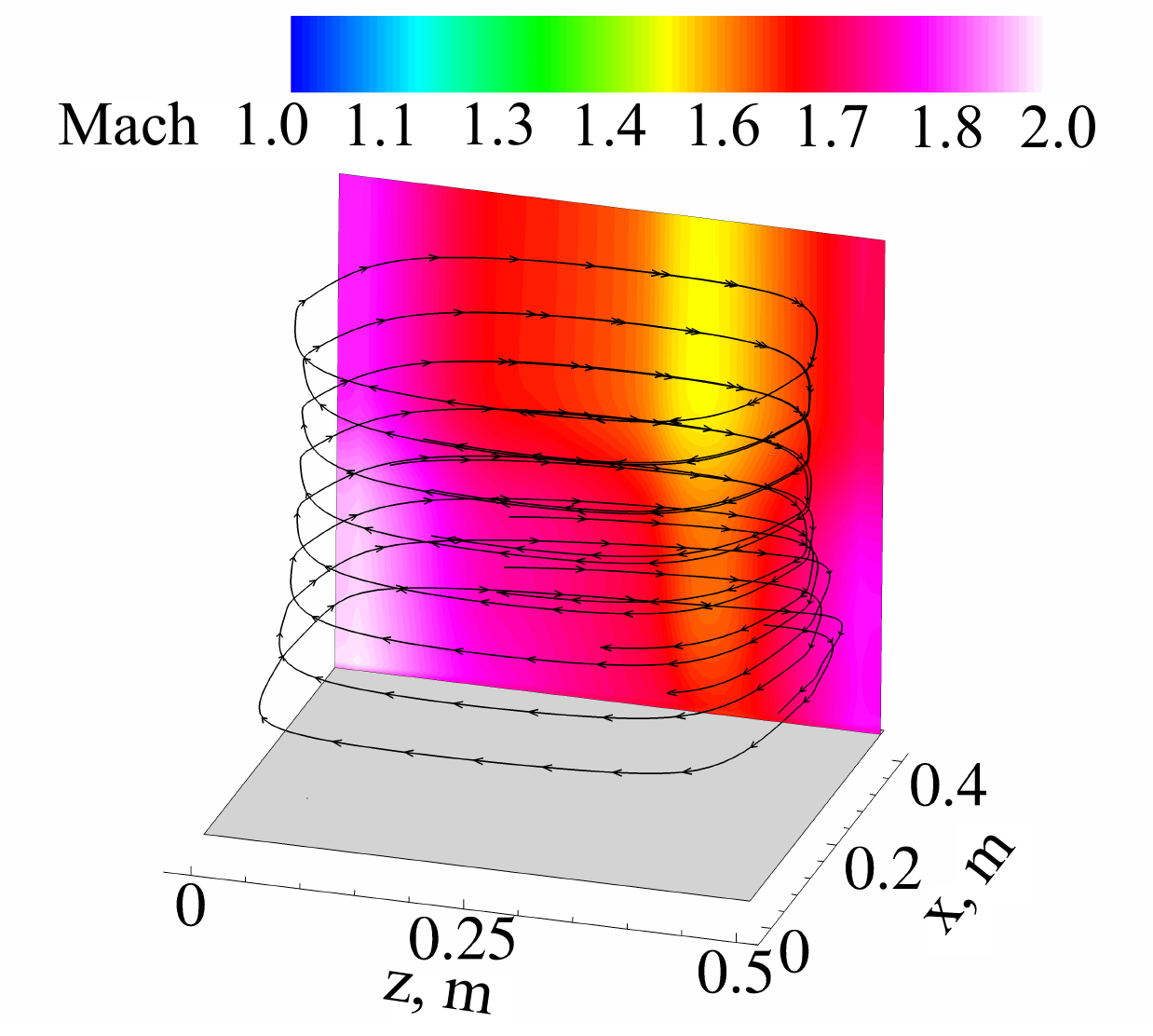}}
     \subfigure[]{\includegraphics[width=0.41\textwidth]{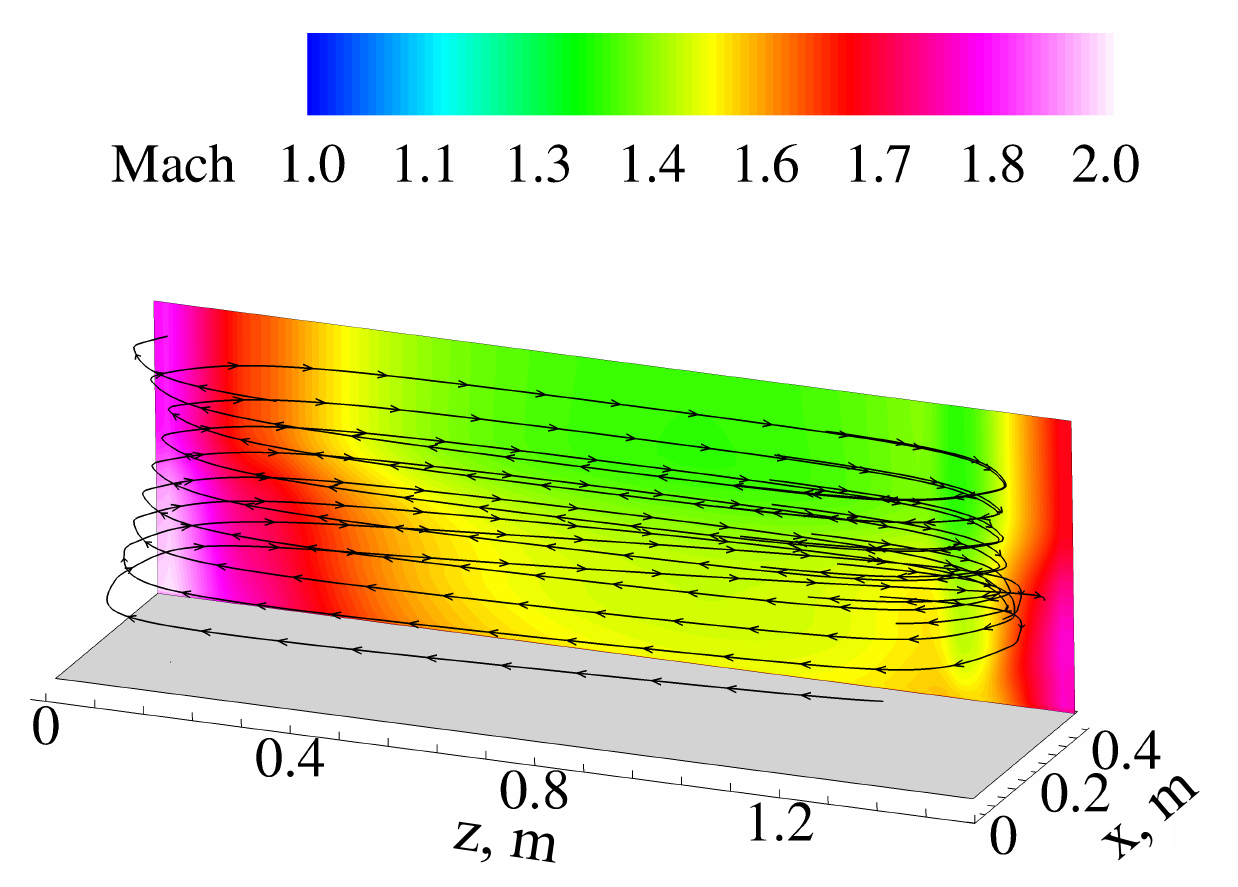}}
     \figurecaption{Current density streamlines and Mach number contours for the electrodeless configuration with $L_{\rm s}=1$~cm for (a) $D=0.5$~m, and  (b) $D=1.5$~m.}
     \label{fig:mach_current_electrodeless}
\end{figure}

The domain's depth significantly influences the Lorentz forces. As seen in Table \ref{tab:F_vs_domain_depth}, when the depth triples, the Lorentz force in the streamwise direction increases by up to 8 times. To understand why this is happening, we can derive a simplified theoretical model staring from the Faraday EMF Eq. (\ref{eqn:FaradayEMF}). Then, the current density along the $z$-axis can be expressed as follows:
\begin{equation}
 J_z = \sigma V_x B_y
\end{equation}
with $\sigma$ the electrical conductivity. The Lorentz force simply corresponds to
\begin{equation}
 F_{x} = J_z B_y \mathrlap{-}V = \sigma V_x B_y^2 \mathrlap{-}V
\end{equation}
with $\mathrlap{-}V$ the volume of the flow subject to a magnetic field strength $B_y$. Thus, clearly, because the volume is proportional to the depth of the domain, we would expect the Lorentz force to scale linearly with the domain depth.
Then, why are we noticing here a different trend showing a much more than expected impact on the force by the domain depth? 

The stronger impact of domain depth on Lorentz forces in our fully-coupled MHD simulations can be attributed primarily to a 3D effect that is not present within the simple analytical prediction. As the current flows between electrodes, it doesn't uniformly permeate the entire MHD region. Instead, it tends to concentrate near the aeroshell surface where the electrodes are placed. This tendency is accentuated when the electrode distance (which is proportional to domain depth) is small. Consequently, in cases of shallow domain depth, the majority of the current remains near the aeroshell surface and this results in a lower Lorentz force compared to analytical predictions. On the other hand, with increased domain depth, the current spreads out across a larger area of the flow and this results in a more substantial Lorentz force which aligns more closely with analytical theory.

The regions where the Lorentz force is most pronounced can be best identified by observing the variations in Mach number contours. Because the Mach number  depends on both flow velocity and sound speed it is a good measure of the Lorentz force (because the latter affects the flow speed) and of the  Joule heating (because the latter affects the flow temperature and speed of sound). These two effects contribute to a decrease in the Mach number. The Lorentz force leads to deceleration and a subsequent reduction in the Mach number because it acts opposite to the flow direction. Similarly, Joule heating also decreases the Mach number by raising the flow temperature, and consequently, the speed of sound. Thus, changes in Mach number serve as a reliable gauge of the strength of MHD effects. In Figure \ref{fig:mach_current_twoelectrode}, the Mach number contours at the domain exit clearly show where the Lorentz force is most pronounced. When the distance between electrodes increases, a significant portion of the flow experiences a Mach number reduction of 15\% or more. With shorter electrode distances, however, only about one third of the flow encounters such substantial changes because the current tends to concentrate near the aeroshell surface rather than spread uniformly across the MHD domain.

It is noteworthy that a similar pattern emerges in the electrodeless configuration. As with the two-electrode configuration, increasing domain depth results in a significantly greater increase in the Lorentz force than is expected from basic theoretical analysis. A simplified theoretical model which assumes uniform current distribution within each magnet domain shows a linear scaling of the Lorentz force with domain depth. However, this is not the case in our numerical results. As outlined in Table \ref{tab:F_vs_domain_depth}, the Lorentz force increases by a factor of 7 when the domain depth triples. Such is nearly twice as much as predicted by basic MHD theory. Unlike the two-electrode configuration, the discrepancy here cannot be attributed to the current's concentration near the aeroshell surface with a shorter domain depth. As evidenced by the current density streamlines in Fig.\ \ref{fig:mach_current_electrodeless}, the current is well-distributed throughout the MHD domain. The reason for the discrepancy between analytical and numerical results lies in the assumption made by the analytical model that the current solely points in one direction within each magnet domain: perpendicular to both the magnetic field vector and the plasma flow velocity. However, in our fully coupled MHD simulations, we find that this assumption does not hold true. A significant portion of the current cannot be perpendicular to both the magnetic field and the flow velocity because it must form a closed loop from one magnet domain to another (refer to current streamlines in Fig.~\ref{fig:mach_current_electrodeless}). This issue is mitigated when the domain depth increases because the current is more aligned within each magnet domain. The advantage of a larger domain depth in this context arises from the fact that a smaller proportion of the domain is required for the current to change direction.

\subsection{Effect of Magnet Separation Distance}

\begin{table}[t]
  \center\fontsizetable
  \begin{threeparttable}
    \tablecaption{Effect of magnet separation distance on the MHD efficiency\tnote{(a)}}
    \label{tab:F_vs_Lt}
    \fontsizetable
\begin{tabular*}{0.48\textwidth}{c@{\extracolsep{\fill}}ccc}
\toprule
$L_{\rm s}$, m& $F_{x}$, kN& $\eta_{\rm MHD}$ & $\eta_{\rm MHD,~theory}$    \\
\midrule
$0.01$  &  $-16.55$  & 0.93 & 0.998\\
$0.50$  &  $-17.16$  &  0.97  & 0.92\\
$2.00$  &  $-10.76$   &  0.61 &  0.74\\
\bottomrule
\end{tabular*}
\begin{tablenotes}
\item[{a}] The domain depth $D=1.5$~m.
\end{tablenotes}
   \end{threeparttable}
\end{table}

\begin{figure}[ht!]
     \centering
     \subfigure[]{\includegraphics[width=0.36\textwidth]{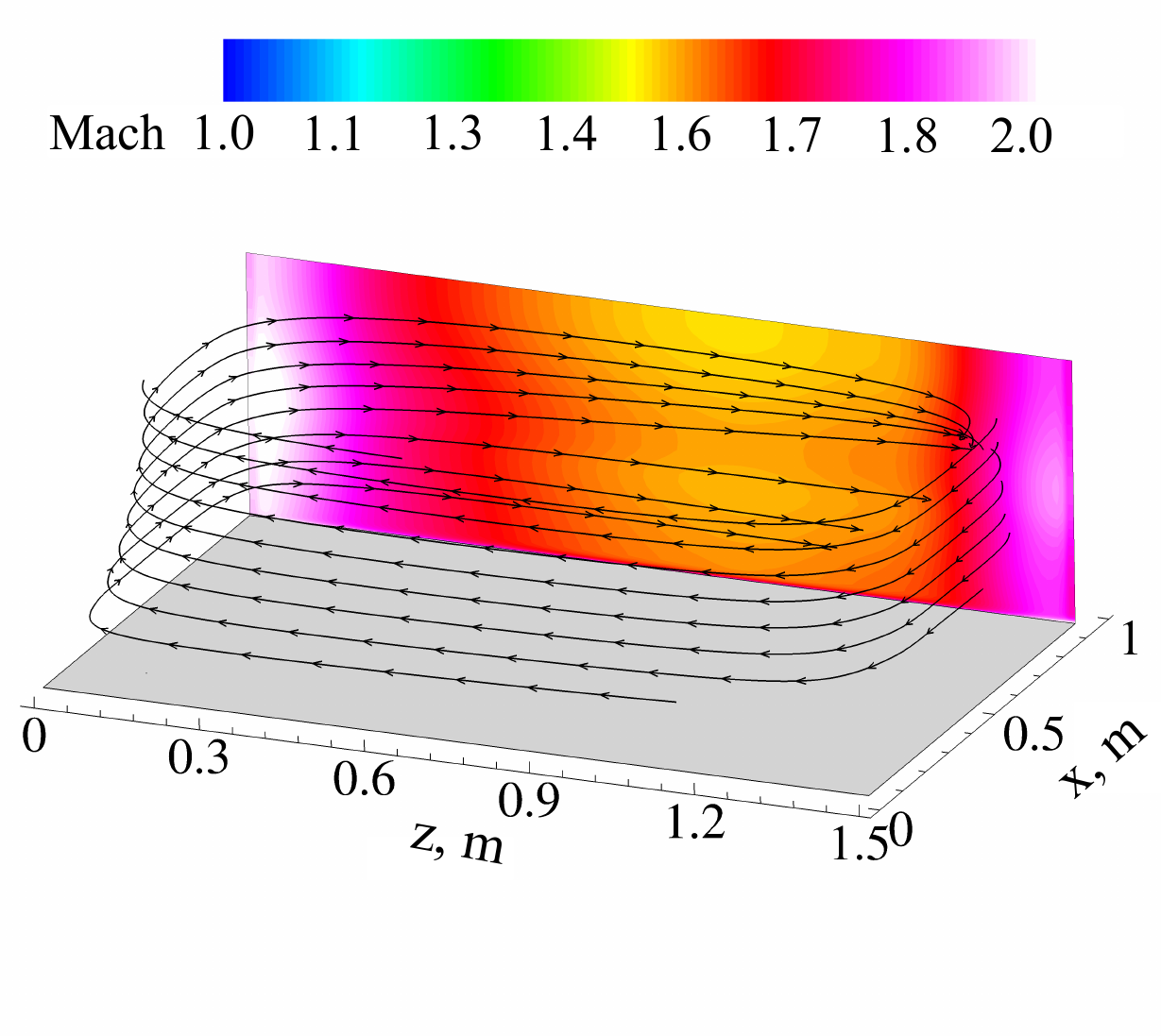}}
     \subfigure[]{\includegraphics[width=0.36\textwidth]{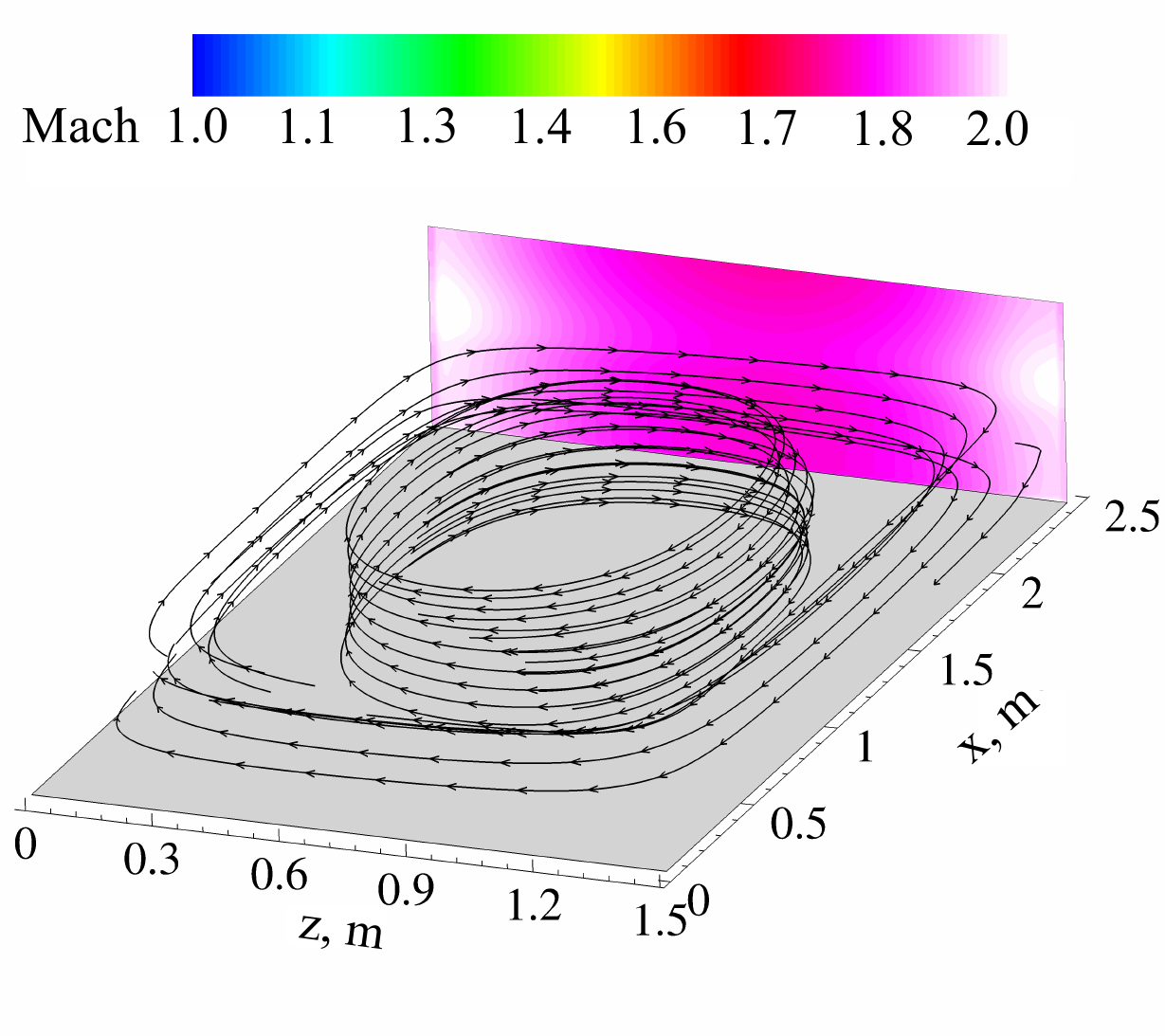}}
     \figurecaption{Current density streamlines and Mach number contours for the electrodeless configuration for (a) $L_{\rm s}=0.5$~m, and  (b) $L_{\rm s}=2.0$~m.}
     \label{fig:magnet_current_electrodeless}
\end{figure}

So far, our evaluation has focused on the performance of an electrodeless two-magnet MHD system wherein the magnetic field of the first magnet transitions over a short span of 1 cm to the magnetic field of the second magnet. This rapid changeover could be challenging to execute in practical applications because the two magnetic fields point in opposite directions. A more practical scenario involves the magnetic field altering its direction over a greater distance that is comparable to or longer than the length of each magnet domain. Consequently, we here investigate the impact of the magnet separation distance on the performance of the electrodeless system while maintaining all other parameters constant.

Table~\ref{tab:F_vs_Lt} outlines a parametric investigation into the effect of the magnet separation distance (denoted as $L_{\rm s}$) keeping the domain depth constant at 1.5~m. It is evident that there is minimal change in the net force  when the magnet separation distance is maintained at or below 0.5~m. Conversely, once the separation distance reaches 2~m the net force experiences a significant decrease. This decrease is not unexpected and can be attributed to the heightened resistance encountered by the current as the distance between the magnets increases. To estimate the amount of voltage lost between the two magnet zones, we first define $\Delta \phi$ as the voltage difference experienced by the current when flowing (along the $x$ direction) from one magnet zone to the other. Thus, the current density between the two magnets can be expressed as:
\begin{equation}
J_x  = \sigma \frac{\Delta \phi}{L_{\rm s}}
\end{equation}
with $\sigma$ representing the conductivity and $L_{\rm s}$ denoting the magnet separation distance. It is noted that the current is equivalent to the product of the current density and the cross-sectional area. In this context, the cross-sectional area corresponds to the height of the plasma multiplied by half the depth because the current flows downstream in the first half of the domain from the first to the second magnet, and upstream in the second half of the domain. Consequently, the current flowing in the MHD system can be expressed as $I=J_x \frac{D}{2} H$, or:
\begin{equation}
I=\sigma \frac{\Delta \phi}{L_{\rm s}} \frac{D}{2} H
\label{eqn:I1}
\end{equation}
Let us temporarily set this aside. It is noted that the current, which flows perpendicular to the flow within each magnet zone and is induced by the Faraday EMF, is diminished by the voltage difference $\Delta \phi$ because the latter operates in a direction opposite to the Faraday EMF. Thus, the overall current density within each magnet zone is:
\begin{equation}
J_z=\sigma \frac{\mathcal{E}-\Delta \phi}{D}
\end{equation}
We can find the total current by multiplying the current density by the cross sectional area which, within each magnet zone, corresponds to the height of the plasma flow $H$ times the length of each magnet zone along $x$, $L_{\rm m}$:
\begin{equation}
I=\sigma L_{\rm m} H \frac{\mathcal{E}-\Delta \phi}{D} 
\label{eqn:I2}
\end{equation}
Because the current flowing within each magnet zone must equal the current flowing from one magnet zone to the other, we can equate Eq.~(\ref{eqn:I1}) and Eq.~(\ref{eqn:I2}). After doing so and isolating the ratio of the voltage drop between magnet zones and the Faraday EMF, we obtain the following expression:
\begin{equation}
 \frac{\Delta \phi}{\mathcal{E}}  =  \left(1+\frac{D^2}{2L_{\rm m} L_{\rm s} }\right)^{-1}
 \label{eqn:DeltaVoverEMF}
\end{equation}
The latter non-dimensional parameter is directly related to the drop in the Lorentz force. Specifically, the voltage drop between magnet zones $\Delta \phi$ also corresponds to the voltage drop within each magnet zone. Thus, the voltage drop $\Delta \phi$ can be perceived as a ``nullifier'' of the Faraday EMF in this system. When the magnitude of the voltage drop approaches the Faraday EMF, the flow of current within each magnet zone is significantly curtailed and, therefore, the Lorentz force goes towards zero. 
\begin{figure}[!t]
     \centering
     \subfigure[]{\includegraphics[width=0.34\textwidth]{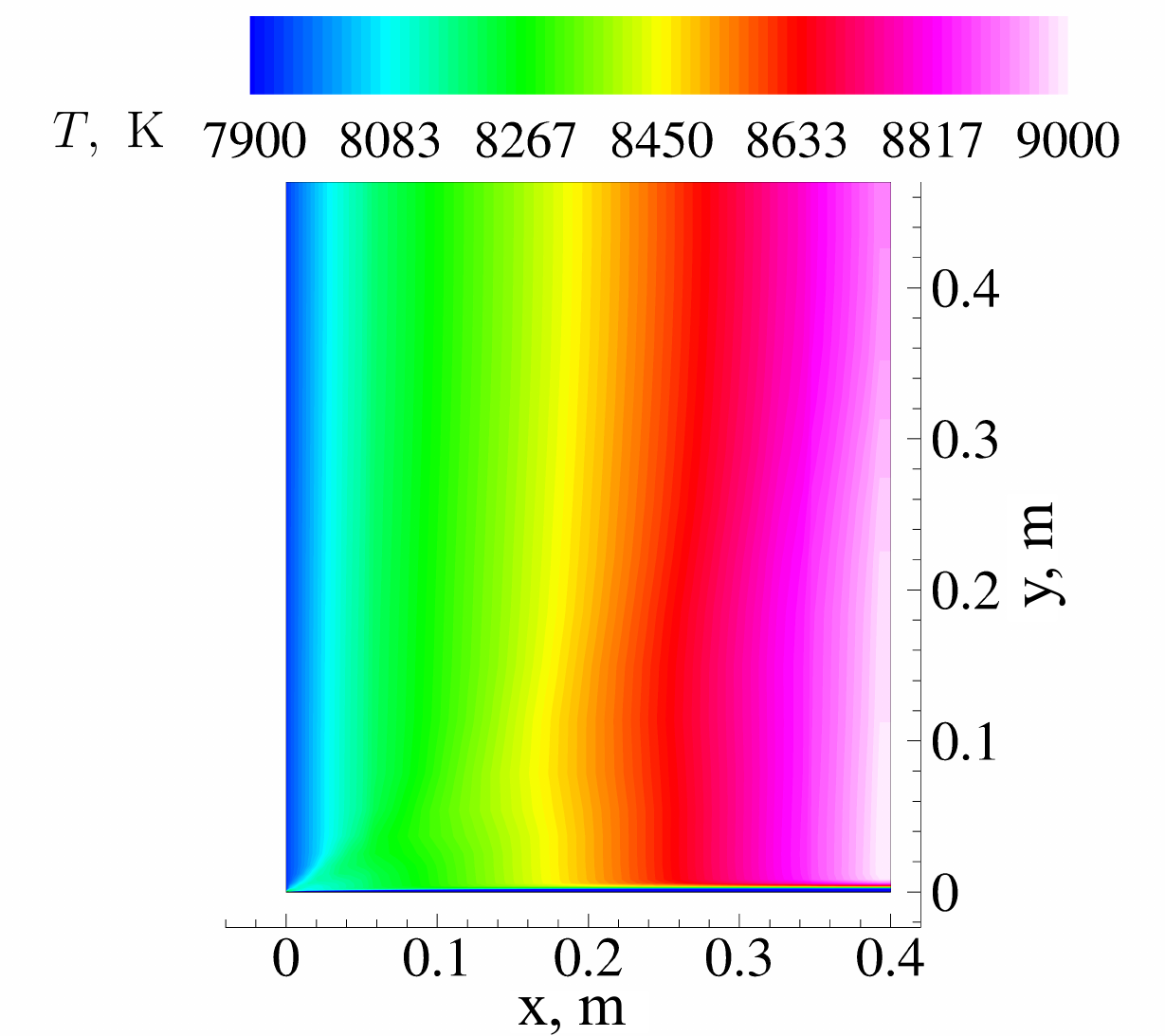}}
     \subfigure[]{\includegraphics[width=0.34\textwidth]{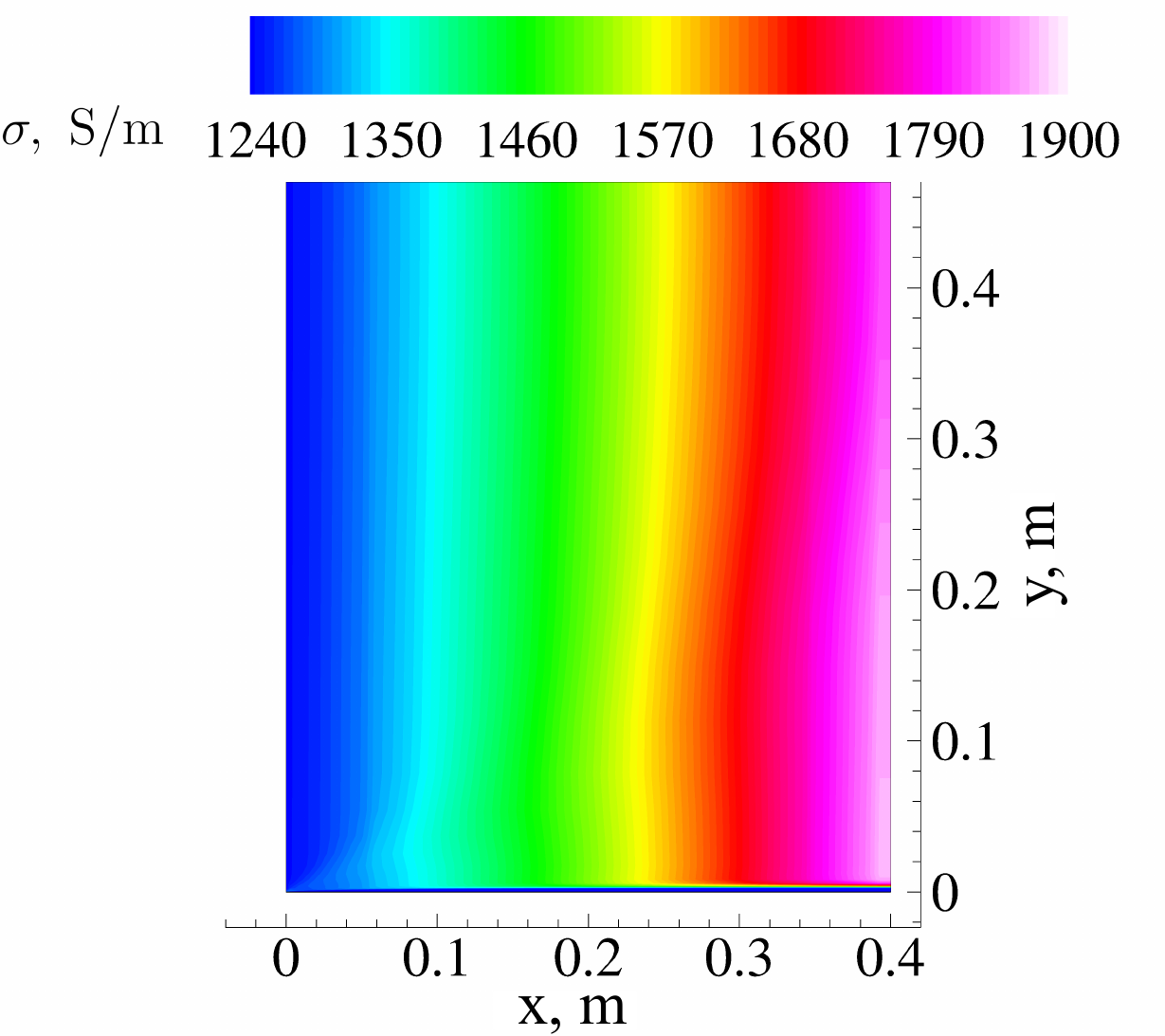}}
     \figurecaption{Contours at the center plane between electrodes with $\gamma_e=100$ of (a) gas temperature and (b) electrical conductivity.}
     \label{fig:sigma_t_contours}
\end{figure}

We can also utilize the aforementioned dimensionless parameter to derive a theoretical equation for the efficiency of our MHD system. First, we define the MHD force generating system efficiency as the ratio between the Lorentz force that is obtained in practice and the one  obtained in ideal conditions (with no voltage drop between magnet zones and with the current density vector having the same magnitude and flowing perpendicularly to the flow at all locations). We then obtain the following
\begin{equation}
\eta_{\rm MHD} \equiv \frac{F_{\rm actual}}{F_{\rm ideal}}
\end{equation}
with the ideal Lorentz force corresponding for the 2-magnet system to:
\begin{equation}
F_{\rm ideal}=J_z B_y 2 D H L_{\rm m}
\end{equation}
In ideal conditions, the current density $J_z$ can be obtained from the Faraday EMF as $J_z=\sigma \mathcal{E}/D$. Also, recall that the Faraday EMF is $\mathcal{E}=V_x D B_y$. After substituting the current density and Faraday EMF the ideal Lorentz force becomes: 
\begin{equation}
F_{\rm ideal}=\sigma V_x  B_y^2 2 D H L_{\rm m}
\label{eqn:Florentzideal}
\end{equation}
We can then find a theoretical approximation to the real Lorentz force in a similar manner but by subtracting $\Delta \phi$ from the Faraday EMF. This yields a current density of $J_z=\sigma (\mathcal{E}-\Delta \phi)/D$. Then we obtain:
\begin{equation}
F=\sigma \mathcal{E} \frac{1-\frac{\Delta \phi}{\mathcal{E}}}{D} B_y 2 D H L_{\rm m}
\end{equation}
After substituting $\mathcal{E}$ from Eq.~(\ref{eqn:FaradayEMF}) and $\Delta \phi / \mathcal{E}$ from Eq.~(\ref{eqn:DeltaVoverEMF}) and simplifying:
\begin{equation}
F=2  H L_{\rm m} \sigma V_x D B_y^2 \left(1-\left(1+\frac{D^2}{2L_{\rm m} L_{\rm s} }\right)^{-1}\right) 
\label{eqn:Florentztheory}
\end{equation}
Then, we can find a theoretical expression for the MHD system efficiency by dividing the theoretical Lorentz force shown in the latter by the ideal Lorentz force outlined in (\ref{eqn:Florentzideal}). After some simplifications, the following expression is obtained:
\begin{equation}
\eta_{\rm MHD,~theory}=1-\left(1+\frac{D^2}{2L_{\rm m} L_{\rm s} }\right)^{-1}
\label{eqn:etaMHDtheory}
\end{equation}
The latter equation serves as a reasonably accurate estimation of the MHD system's efficiency, as corroborated by the outcomes presented in Table \ref{tab:F_vs_Lt}, where a good agreement is observed between the former and the efficiency derived from numerical outcomes.  The discrepancy between the numerically-obtained efficiency and the theoretical prediction is attributed to the current not travelling  perpendicular to the flow when entering and leaving each magnet zone. This effect is more pronounced when the magnet separation distance is larger (see Fig.~\ref{fig:magnet_current_electrodeless}), thus leading to a slightly greater discrepancy with the theoretical prediction (which assumes that the current travels perpendicular to the flow within each magnet zone).  Another source of error associated with  Eq.~(\ref{eqn:etaMHDtheory}) is that it assumes that the electrical conductivity is constant within the whole domain. This is not the case because, when the Lorentz force opposing the flow velocity is significant, the MHD process converts the kinetic energy of the flow into thermal energy effectively raising its temperature (see Fig.~\ref{fig:sigma_t_contours}a). This leads to a significant increase in conductivity because, when the plasma is strongly ionized (as is the case here), the electrical conductivity increases proportionally to the temperature to the $\frac{3}{2}$ power (see Fig.~\ref{fig:sigma_t_contours}b). This is a possible cause to the MHD efficiency obtained numerically being slightly higher than the theoretical expression in some cases.

Interestingly, Eq.~(\ref{eqn:DeltaVoverEMF}) illustrates that as the magnet separation distance becomes very large, the reduction in the Lorentz force is proportional to $(2L_{\rm m} L_{\rm s} )/D^2$. This implies that a greater domain depth not only enhances the efficiency of the MHD process by aligning the current streamlines more perpendicularly to the flow (as discussed in the previous subsection), but it also significantly aids in minimizing the voltage drop between magnets (and consequently, in limiting the decrease in the Lorentz force) when the separation distance between the magnets is substantial.

\subsection{Effect of Induced Magnetic Field}

\begin{figure*}[!t]
     \centering
     \subfigure[$z=0.1D$]{\includegraphics[width=0.26\textwidth]{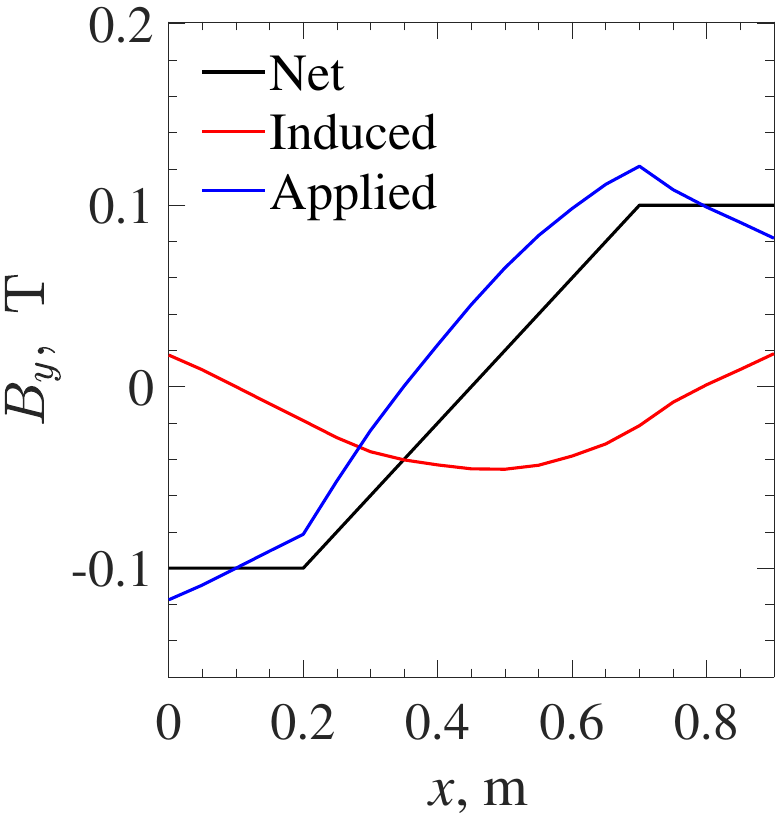}} ~~~~~~~~~
     \subfigure[$z=0.5D$]{\includegraphics[width=0.26\textwidth]{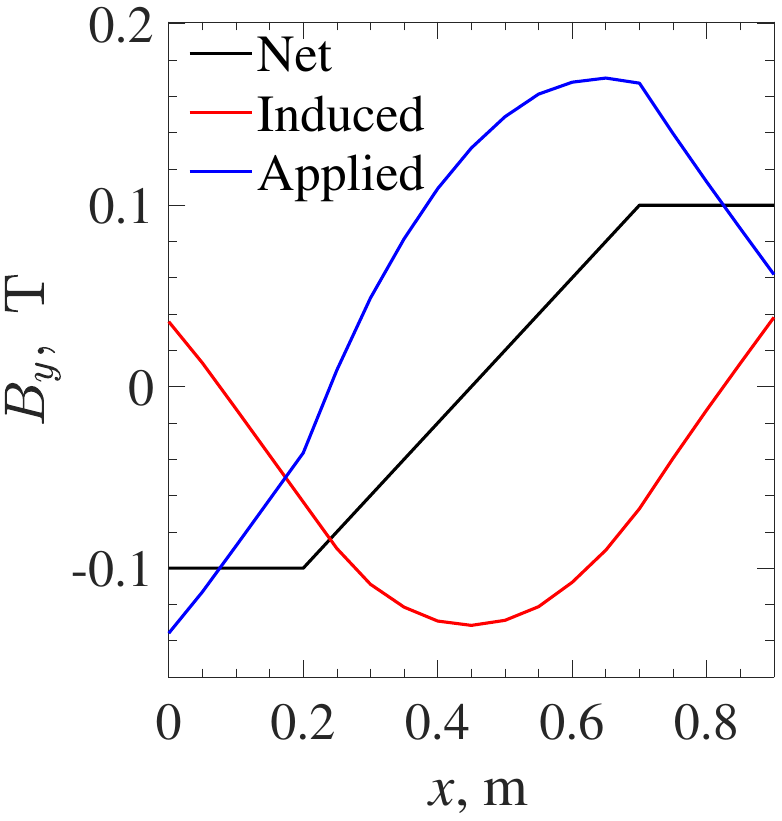}} ~~~~~~~~~
     \subfigure[$z=0.9D$]{\includegraphics[width=0.26\textwidth]{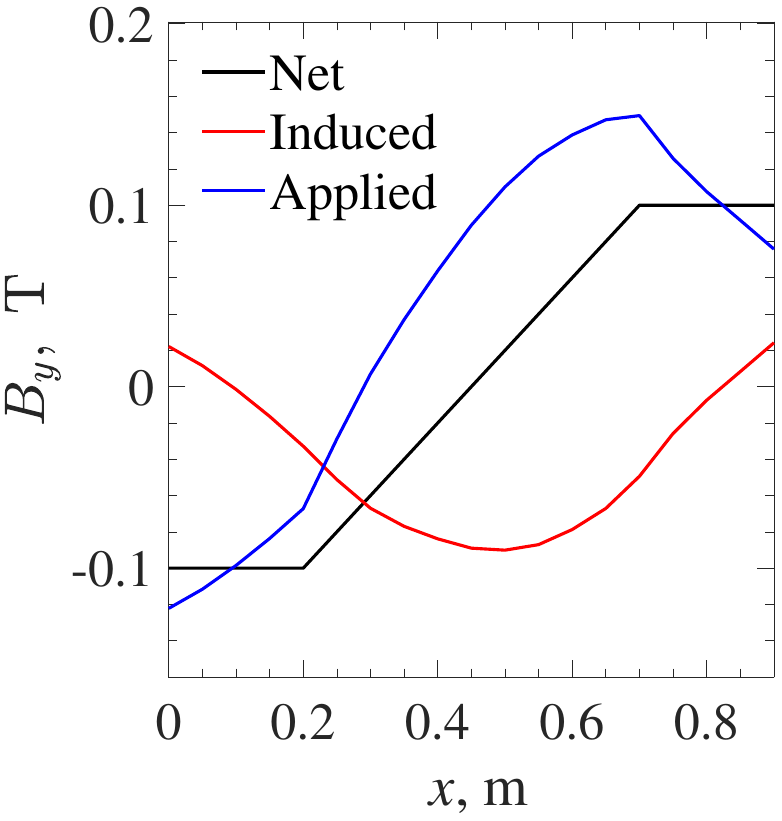}}     
     \figurecaption{Comparison between the induced and applied magnetic fields for the case where $D=1.5$ m and the separation between magnets is $L_{\rm s}=0.5$ m.}
     \label{fig:inducedmagfield}
\end{figure*}

{
 For the electrodeless MHD patch proposed herein, the current flows in a loop above the surface in a manner similar to how the current flows in a solenoid. Because the current within the MHD patch tends to be large (50 kA and more), it could therefore lead to an induced magnetic field  that is substantial. This is a source of concern because the physical model used in this study does not include transport equations to determine the induced magnetic field. Thus, we here aim to assess the impact of the induced magnetic field on the system. We can find precisely the induced magnetic field at a single location through the use of the Biot-Savart law \cite[p.~138]{book:2013:grant} integrated over all the cells in the domain:
\begin{equation}
\vec{B}_{\rm induced}=\frac{\mu_0}{4 \pi} \int_{\mathrlap{-}V} 
\frac{\vec{J} \times \vec{r}}
     {\left|\vec{r}\right|^3} d{\mathrlap{-}V}
\end{equation}
with $\vec{J}$ the current density in the cell, and $\vec{r}$ the vector difference between the point where the induced magnetic field is evaluated and the position of the cell of volume $d\mathrlap{-}V$ where the induced magnetic field is produced.}

{As can be seen in Fig.~\ref{fig:inducedmagfield}, the induced magnetic field points primarily from the plasma towards the aeroshell and its magnitude reaches a maximum value of 0.12~T in the center of the current loop. Such is of the same order of magnitude as the applied magnetic field generated by each magnet. It can thus be concluded that the induced magnetic field does play a significant role for these problems and can not be ignored. The results obtained in the previous sections are not all lost, however. They remain accurate as long as the magnetic field used in the Physical Model (Section II) and specified in the Problem Setup (Section IV) is understood to be the \emph{net magnetic field}. The net magnetic field is defined as the sum of the applied magnetic field originating from the magnets and the induced magnetic field originating from the current in the plasma:
\begin{equation}
\vec{B}=\vec{B}_{\rm applied} + \vec{B}_{\rm induced}
\end{equation}
Therefore, in order to obtain the same Lorentz forces as found previously but taking into consideration the effect of the induced magnetic field, it follows that the applied magnetic field needs to be corrected as $\vec{B}_{\rm applied}= \vec{B}- \vec{B}_{\rm induced}$. As shown in Fig.~\ref{fig:inducedmagfield}, this leads to a small change in the applied magnetic field originating from the magnets. The most notable difference is that the peak applied magnetic field is now of 0.16~T instead of 0.1~T. It is emphasized that should the code include additional transport equations that solve for the induced magnetic field, the same current distribution, Mach number distribution, and Lorentz forces would be obtained as in the previous subsections as long as the magnets are arranged such that the sum of their magnetic fields corresponds to the ``applied magnetic field'' depicted in Fig.~\ref{fig:inducedmagfield}.  }

{
The magnetic field induced by the large currents within the plasma layer will not alter the currents flowing within the electromagnets unless the plasma currents become unstable and exhibit some variations in time (which is not the case here). However, because the induced magnetic field points primarily in the negative $y$ direction within each magnet region (see Fig.~\ref{fig:inducedmagfield}), such could lead to an additional torque on the system composed of the two electromagnets. Indeed, the first electromagnet magnetic field points in the same direction as the induced magnetic field thus leading to the first magnet feeling a force directed towards the plasma layer. Because it is oriented in the opposite direction, the second magnet feels a force directed away from the plasma layer. Therefore, the induced magnetic field in the plasma layer is expected to lead to a net torque on the magnet system.}

\section{Conclusions}

A novel MHD system tailored for aerocapture is introduced herein. The new design offers distinct advantages over its predecessors by eliminating the need for electrodes and by operating in a localized manner. This allows the MHD domain to be confined to a small region of flow (as opposed to encompassing the entire flow around the capsule), thereby reducing magnet mass and facilitating localized lift and drag forces. Moreover, the absence of electrodes mitigates numerous issues commonly encountered in multi-electrode MHD systems such as  electrode oxidation and erosion,  electrode cooling, and the cooling of the external circuit due to the substantial currents flowing between electrodes.

The performance of the two-electrode MHD system is noticeably influenced by the non-neutral cathode sheath. Given that thermionic emission is unlikely to be significant in Earth entry flows due to electrode oxidation, the sole mechanism for reducing the resistivity of the sheath is through secondary electron emission. Even with a significantly high secondary electron emission (SEE) coefficient of 10, it is here seen that the plasma sheath diminishes the Lorentz force by more than 10 times.

The electrodeless system consistently generates Lorentz forces that are several times higher than those of the two-electrode system when using the same magnetic field strength. The two-electrode system's performance only approaches the electrodeless system's when the effective SEE coefficient is raised to an improbable value of 100. The higher efficiency of the electrodeless system is attributed to the lack of current losses caused by cathode sheath resistance and to the prevention of current accumulation near the aeroshell surface.

\appendix
\section{Vibrational-translational energy exchange relaxation distance}

 {
The vibrational-translational (v--t) energy relaxation distance can be expressed as the product between the flow speed and the v--t energy relaxation time. From Ref.~\cite{jpp:2007:parent}, an expression for the v--t relaxation time can be obtained from $1/\tau_{\rm v-t} \sim 7\cdot 10^{-16}\cdot N \cdot \exp(-141 \cdot T^{-1/3})$. For high efficiency, the MHD region needs to be located where the Faraday EMF is high, i.e., where the product between the flow velocity and the electrical conductivity is high. Such occurs where the local Mach number is approximately 2. Noting that $V\approx M \sqrt{\gamma R T} \approx 2\sqrt{\gamma R T}$, and using the ideal gas law $N=P/(k_{\rm B} T)$, the relaxation distance  then becomes:
\begin{equation}
L_{\rm v-t} \approx \frac{2 k_{\rm B} T^{1.5}  \sqrt{\gamma R } }{ 7\cdot 10^{-16} \cdot  P \cdot \exp(-141 \cdot T^{-1/3})}
\end{equation}
Because most of the flow entering the MHD region is compressed through an oblique shock with a turning angle of about 35 degrees, it can be easily shown that the pressure at the entrance of the MHD region is about half the stagnation pressure after the normal shock on the stagnation line more or less independently of flight Mach number in the range $25 < M_\infty < 40$. Further, at hypersonic speeds, the stagnation pressure on the stagnation line is very close to 2 times the flight dynamic pressure. In other words, the pressure at the entrance of the MHD region is approximately the same as the flight dynamic pressure, independently of altitude or flight Mach number:
\begin{equation}
L_{\rm v-t} \approx \frac{2 k_{\rm B} T^{1.5}  \sqrt{\gamma R } }{7\cdot 10^{-16}\cdot  P_{\rm dyn} \cdot \exp(-141 \cdot T^{-1/3})}
\end{equation}
For $\gamma\sim 1.4$ and $R\sim 600$~J/kgK, and knowing that $T$ at the entrance of the MHD region would vary between 5000~K and 8000~K between the flight Mach numbers of 25 and 40, we arrive at the following expression (accurate to within 25\%):
\begin{equation}
L_{\rm v-t} \approx \frac{1200 ~\textrm{Pa-m}}{P_{\rm dyn}}
\label{eqn:Lvt}
\end{equation}
Because the relaxation distance depends only on dynamic pressure, and because the dynamic pressure depends only on flight Mach number and altitude, the latter expression can be used to find the relaxation distance given altitude and flight Mach number.}

\section{Electron-translational energy exchange relaxation distance}

{
When the plasma has an ionization fraction greater than $10^{-4}$  or so (as is the case here), the relaxation time of the free electron energy corresponds approximately to the ratio between the electron energy and the  electron-ion energy relaxation term. After substituting the electron-ion energy relaxation term from Ref.~\cite[Eq.~21]{pof:2024:parent} we obtain:
\begin{equation}
\tau_{\rm e-t}\approx \left(\frac{3}{2} N_{\rm e} k_{\rm B} T_{\rm e}\right) \left/ \left(\frac{ N_{\rm e}  N_{\rm i} T_{\rm e}  6 k_{\rm B} C_{\rm e}^4 \ln \Lambda}{\pi^3 \epsilon_0^2 m_{\rm e}m_{\rm i} \overline{q_{\rm e}}^3 }\right.\right)
\end{equation}
Consider a quasi-neutral plasma where $N_{\rm i}\approx N_{\rm e}$.
Also note that the electron-translational (e--t) energy relaxation distance corresponds to the local flow velocity times the e--t relaxation time. Because the Mach number at the entrance of the MHD region is about 2, we can then say:
\begin{equation}
L_{\rm e-t} \approx \frac{\sqrt{\gamma R T} \pi^3 \epsilon_0^2 m_{\rm e}m_{\rm i} \overline{q_{\rm e}}^3 }{2 N_{\rm e} C_{\rm e}^4 \ln \Lambda}
\end{equation}
We can further divide and multiply the denominator by $N$  and use the ideal gas law such that $N=P/(k_{\rm B}T)$. Then we obtain:
\begin{equation}
L_{\rm e-t} \approx \frac{\sqrt{\gamma R} \pi^3 \epsilon_0^2 m_{\rm e}m_{\rm i} \overline{q_{\rm e}}^3 k_{\rm B}T^\frac{3}{2}}{2 P \frac{N_{\rm e}}{N} C_{\rm e}^4 \ln \Lambda}
\end{equation}
Substitute the electron thermal velocity, consider the case where $T_{\rm e}$ is close to $T$, and recall that $P\approx P_{\rm dyn}$ at the entrance of the MHD region. After some simplifications, we obtain:
\begin{equation}
L_{\rm e-t} \approx \frac{\sqrt{\gamma R} \pi^3 \epsilon_0^2 m_{\rm e}m_{\rm i}  k_{\rm B}T^3 }{2 P_{\rm dyn} \frac{N_{\rm e}}{N} C_{\rm e}^4 \ln \Lambda} \left(\frac{8 k_{\rm B}}{\pi m_{\rm e}} \right)^\frac{3}{2}
\end{equation}
We can simplify the latter greatly by setting $\gamma=1.4$, $R=600$~J/kgK, by setting the ion mass to the one of the atomic nitrogen ion, and by setting the Coulomb logarithm $\ln \Lambda$ to 6.  Also, we note that the temperature at the entrance of the MHD region varies between 6000 and 8000~K for a flight Mach number range 25--40. Then, we obtain the following approximate expression (accurate within 30\%):
\begin{equation}
L_{\rm e-t} \approx \frac{0.2~\textrm{Pa-m} }{\frac{N_{\rm e}}{N} \cdot P_{\rm dyn}  } 
\label{eqn:Let}
\end{equation}
}

\bibliography{all}
\bibliographystyle{aiaa2}

\end{document}

%use the following template for figure with subfigures:
\begin{figure}[ht!]
     \centering
     \subfigure[(a) blah blah]{\includegraphics[width=0.49\textwidth]{figs/fig1.pdf}}
     \subfigure[(b) blah blah]{\includegraphics[width=0.49\textwidth]{figs/fig2.pdf}}
     \figurecaption{Caption here.}
     \label{fig:name}
\end{figure}

% use the following template for a table:
\begin{table}[ht]
    \centering
    \tablecaption{Air plasma frequency as a function of electron number density.}
    \begin{tabular*}{\textwidth}{ll}
    \toprule
    Electron number density, m$^{-3}$ & Plasma frequency, MHz\\
    \midrule    
    $10^{12}$ & 9 \\
    $10^{13}$ & 28 \\
    \bottomrule
    \end{tabular*}
    \label{tab:plasmafrequency}
\end{table}

\begin{table*}
\centering
\tablecaption{Effect of grid size of the 2D axisymmetric simulation on the electron losses and gains (in kg/s) in the complete flow domain $0\leq x/R \leq 8.21$. The simulations here represent flow around RAM-C-II at 71 km altitude conditions simulated using the Dixon-Lewis transport model and the Kim chemical model.}
\begin{tabular}{lccc}
\toprule
\multirow{2}*{Electron gain/loss mechanism} &  \multicolumn{3}{c}{grid size} \\
& $250\times180$ & $500\times360$ & $1000\times720$  \\
\midrule
Convection through outflow boundary & 	-5.525$\times 10^{-11}$ & 	-5.290$\times 10^{-11}$ & 	-5.190$\times 10^{-11}$ \\ 
Surface catalyticity & 	-4.524$\times 10^{-11}$ & 	-4.679$\times 10^{-11}$ & 	-4.735$\times 10^{-11}$ \\ 
2-body recombination within the plasma & 	-1.280$\times 10^{-08}$ & 	-1.235$\times 10^{-08}$ & 	-1.218$\times 10^{-08}$ \\ 
3-body recombination within the plasma & 	-4.827$\times 10^{-15}$ & 	-4.640$\times 10^{-15}$ & 	-4.533$\times 10^{-15}$ \\ 
Electron impact ionization & 	2.716$\times 10^{-12}$ & 	2.391$\times 10^{-12}$ & 	2.317$\times 10^{-12}$ \\ 
Associative ionization & 	1.290$\times 10^{-08}$ & 	1.245$\times 10^{-08}$ & 	1.228$\times 10^{-08}$ \\ 
\bottomrule
\end{tabular}
\label{tab:gainslosses_gcs}
\end{table*}

\begin{figure}[h!]
     \centering
     \subfigure[61 km]{\includegraphics{figures/Ne_K61.pdf}}
     \hfill
     \subfigure[71 km]{\includegraphics{figures/Ne_K71.pdf}}
     \figurecaption{Comparison of electron number density at the locations corresponding to the leading-edge of the electrostatic probe rake, using the 2D axisymmetric simulations.} \label{fig:2DNeBL}
\end{figure}

%\begin{figure}[h!]
%     \centering
%     \subfigure[temperature]{\includegraphics[]{figures/inflowT.pdf}}
%     \hfill
%     \subfigure[molar fractions]{\includegraphics[]{figures/inflowchi.pdf}}
%     \figurecaption{Comparison of (a) temperature profiles and (b) molar fraction profiles predicted by the different transport models at $x\approx1.132$ m in the 2D axisymmetric case for 71 km altitude. The solid lines correspond to solution obtained using the Dixon-Lewis transport coefficients and the dashed lines correspond to the solution obtained using the Gupta-Yos transport coefficients.}
%\end{figure}